%% file: master.tex
\documentclass[12pt]{iopart}
\usepackage{graphicx}
\usepackage{bm}
\usepackage{iopams}
\usepackage{setstack}

\usepackage{color}

\begin{document}
\title[Level 2 large deviation functionals]{Level 2 large deviation functionals for systems with and without detailed balance}
\author{J Hoppenau, D Nickelsen, and A Engel}
\address{Uni Oldenburg}
\ead{johannes.hoppenau@uni-oldenburg.de}

\begin{abstract}
Large deviation functions are an essential tool in the statistics of rare events. Often they can be obtained by contraction from a so-called level 2 large deviation {\em functional} characterizing the empirical density of the underlying stochastic process. For Langevin systems obeying detailed balance, the explicit form of this functional has been known ever since the mathematical work of Donsker and Varadhan. We rederive the Donsker-Varadhan result by using stochastic path-integrals and then generalize it to situations without detailed balance including non-equilibrium steady states. The proper incorporation of the empirical probability flux turns out to be crucial. We elucidate the relation between the  large deviation functional and different notions of entropy production in stochastic thermodynamics and discuss some aspects of the ensuing contractions. Finally, we illustrate our findings with  examples. 
\end{abstract}


\maketitle

\eqnobysec

\renewcommand{\rho}{\varrho}
\newcommand{\pst}{p_\mathrm{st}}
\newcommand{\peq}{p_\mathrm{eq}}
\newcommand{\vst}{\bi v_\mathrm{st}}
\newcommand{\jst}{\bi j_\mathrm{st}}
\newcommand{\vrho}{\bi v_{\!\rho}}
\newcommand{\jrho}{\bi j_{\!\rho}}
\newcommand{\vp}{\bi v_{p}}
\newcommand{\bbmu}{\boldsymbol{\bar\mu}}
\newcommand{\bbnu}{\boldsymbol{\bar\nu}}
\newcommand{\Si}{S_\mathrm{i}}
\newcommand{\Sna}{S_\mathrm{na}}
\newcommand{\Sa}{s_\mathrm{a}}
\newcommand{\dSi}{\dot S_\mathrm{i}}
\newcommand{\dSna}{\dot S_\mathrm{na}}
\newcommand{\dSa}{\dot S_\mathrm{a}}
\newcommand{\dSar}{\dot S_\mathrm{a}[\rc]}
\newcommand{\Ina}{I_\mathrm{na}}
\newcommand{\Ia}{I_\mathrm{a}}
\newcommand{\Imix}{I_\mathrm{mix}}
\newcommand{\xc}{\bi x(\cdot)}
\newcommand{\rc}{\rho(\cdot)}
\newcommand{\mc}{\bmu(\cdot)}
\newcommand{\nc}{\bnu(\cdot)}
\newcommand{\pc}{p(\cdot)}
\newcommand{\todo}[1]{{\small{{\color{red}#1}}}}

\input{intro}

\input{detailed_balance}
\input{ness}
\input{discussion}

\input{example}
\section{Conclusion}
We presented an instructive rederivation of the Donsker-Varadhan result for the level 2 large deviation functional $I[\rc]$ using a path-integral approach for Langevin systems that obey detailed balance. The derivation makes use of a simple relation between the left and right eigenfunctions of the tilted Fokker Planck operator. Due to the lack of such a relation in cases without detailed balance, a generalization of our derivation to NESSs is not immediate. Introducing the empirical current $\bmu$ and considering $I[\rc, \mc]$ instead of $I[\rc]$, however, provides the missing relation between the eigenfunctions and we arrive at a closed form of $I[\rc, \mc]$ which was also found by Maes et al. \cite{Maes2008} along other lines.

In our derivation it was important to note that the definition of the empirical current implies a vanishing divergence, $\bnabla \bmu=0$, except for the starting and end point of the underlying trajectory. This fact constitutes an argument why $I[\rc, \mc]$ is only to be considered for divergence-less $\bmu$.

Just as the irreversible entropy production splits into an adiabatic and non-adiabatic part for systems violating detailed balance, the large deviation functional $I[\rc,\mc]$ can be written as the sum of two non-negative contributions $\Ia[\rc,\mc]$ and $\Ina[\rc]$, in direct correspondence to the respective entropy production rates. The empirical current $\bmu$ accordingly enters the large deviation functional only by the adiabatic part $\Ia[\rc,\mc]$.

Building on the contraction principle, the ``master'' functional $I[\rc,\mc]$ is a starting point to derive all desirable large deviation functionals and functions. Contraction to $I[\rc]$ and making use of the detailed balance condition reproduces the Donsker-Varadhan result and demands a vanishing optimal current $\bbmu$ for all $\rho$. Large deviation functions of entropy productions follow from contractions of $I[\mc]$, the large deviation functional for the empirical current hence is worthy of more attention than having received by the literature. The contraction to $I[\mc]$, however, turns out to be quite involved, but a set of equations to determine by numerical means the large deviation function for the adiabatic entropy production, $J(\Sa)$, has been set up exemplarily. Carrying out the numerics for $J(\Sa)$ and analytical progress in the determination of $I[\mc]$ are subjects for future efforts.

\section{Acknowledgements}
Financial support from DFG under EN278/9-1 is gratefully acknowledged. We thank Marcel Kahlen for many fruitful discussions.

\section*{References}
\bibliographystyle{unsrt}
\bibliography{ldfv2_lit.bib}

\input{appendix}

 
\end{document}

%% file: intro.tex
\section{Introduction}
\label{sec:intro}



Thermodynamic quantities of small systems fluctuate measurably. With the discovery of fluctuation theorems, the probability distributions of work, heat, and entropy became the focus of stochastic thermodynamics, for recent reviews see \cite{Jarzynski2011, Seifert2012}. In particular, the tails of these distributions turn out to be important for the properties of small systems. Therefore, the statistics of rare events received an increase of interest in the past decades.

The mathematical framework to address the statistics of rare events is {\em large deviation theory} \cite{Derrida2007,Touchette2009}. Of particular interest are quantities of the type
\begin{equation}
	\label{eq:A_T}
  a_T[\xc] = \frac{1}{T}\int_0^T \rmd t\, A(\bi x(t))
  ,
\end{equation}
sometimes called Brownian functionals \cite{Majumdar}. Here $\bi x(t)$ denotes a $d$-dimensional stochastic process. If it is ergodic with some stationary probability measure $\pst(\bi x)$, then, for large oberservation times $T$, $a_T$ will converge to its mean value, 
\begin{equation}
  \lim_{T\to\infty}a_T[\xc]=\langle a\rangle_\mathrm{st}=\int\rmd\bi x\,\pst(\bi x)A(\bi x)
  .
\end{equation}
To characterize the deviations of $a_T$ from $\langle a\rangle_\mathrm{st}$, we write the distribution of $a_T$ in the form 
\begin{equation}
  p_T(a) \equiv p(a\!=\!a_T)=\exp\left[-T J(a) + \mathrm o(T)\right]
  ,
\end{equation}
with the {\em large deviation function}
\begin{equation}
  J(a) = -\lim_{T \to \infty}\frac{1}{T} \ln p_T(a)
  .
\end{equation}
If the limit exists, the random variable $a$ is said to satisfy a large deviation principle.

The large deviation function $J(a)$ contains the desired information about the statistics of $a$. First of all, consistency requires $J(\langle a \rangle_\mathrm{st})=0$ and $J(a)\geq0$ for all $a$. Taylor expansion around the minimum of $J(a)$ up to second order yields a Gaussian fluctuation statistics and generalizes  Einstein´s theory of equilibrium fluctuations \cite{LandauLifshitz1958}. In addition, the complete function $J(a)$ characterizes the statistics of exponenially rare events deviating from $\langle a \rangle_\mathrm{st}$. Recent applications of large deviation functions to describe rare events in statistical mechanics can be found in \cite{Lebowitz1999,BodineauDerrida2007,Speck2012,Nemoto2014,Verley2014b,Harbola2014,Barato2015a}.

%
%
%

Different Brownian functionals deriving from the same stochastic process $\bi x(t)$ have different large deviation functions. On the other hand, we may rewrite \eref{eq:A_T} as
\begin{eqnarray}
  a_T &= \frac{1}{T}\int_0^T\rmd t\, A(\bi x(t)) \int\rmd\bi y\,\delta\left(\bi y-\bi x(t)\right) \nonumber\\
	&= \int \rmd\bi y\, A(\bi y)\,\rho_T(\bi y;\bi x(t)) \label{eq:aT_rho}
\end{eqnarray}
with the so-called empirical density
\begin{equation}
  \rho_T(\bi y;\bi x(\cdot)) = \frac{1}{T} \int_0^T \rmd t\, \delta(\bi y - \bi x(t)) \label{eq:def_rho}
  .
\end{equation}
Due to its dependence on $\bi x(\cdot)$ the \emph{empirical density} is a random function. Clearly, 
\begin{equation}\label{eq:limrho}
  \lim_{T \to \infty} \rho_T(\bi y;\bi x(\cdot))=\pst(\bi y)\; .
\end{equation}
If we assume that $\rho_T$ by itself obeys a large deviation principle,
\begin{equation}
  \label{eq:LDF_def}
  P_T[\rc] \equiv P[\rc\!=\!\rho_T] = \exp\left[-T I[\rc] + \mathrm o(T)\right]
  ,
\end{equation}
with the large deviation {\em functional}
\begin{equation}
  I[\rc] = -\lim_{T \to \infty}\frac{1}{T} \ln P_T[\rc]
  ,
\end{equation}
all large deviation functions of Brownian functionals $a_T$ maybe derived from $I[\rc]$ by contraction:
\begin{equation}
  J(a) = \underset{\rc\,|\,a=\int\!\rmd\bi y\, A(\bi y)\,\rho(\bi y)}{\min} I[\rc] \label{eq:contr_Ja}
  .
\end{equation}
For this reason $I[\rc]$ is known as \emph{level 2} large deviation functional.

To be more specific, let us consider as underlying stochastic process an overdamped Langevin dynamics
\begin{equation}
	\label{eq:langevin1}
	\dot{\bi{x}} = \bm f (\bi x,t) + \frac{1}{\sqrt \beta}\, \bm \xi(t)
	,
\end{equation}
with a force $\bm f(\bm x,t)$ and white noise $\bm \xi(t)$ with correlation $\langle \xi_i(t) \xi_j(t') \rangle = 2\delta_{ij} \delta(t - t')$. The dynamics may also be defined by the Fokker-Planck equation
\numparts
\begin{eqnarray}
  &\partial_t p(\bi x,t) = -\bnabla\cdot \bi j(\bi x,t) \label{eq:conti_eq} 
  , \\
  &\bi j(\bi x,t) = \bi f(\bi x,t)\,p(\bi x,t) - \frac{1}{\beta}\bnabla p(\bi x,t) \label{eq:def_j}
  ,
\end{eqnarray}
\endnumparts
where $p(\bi x,t)$ is the probability density and $\bi j(\bi x,t)$ the probability current density for the state $\bi x$ at time $t$. 
Accordingly, we get for the stationary state with distribution $\pst$ and current $\jst$ 
\numparts
\begin{eqnarray}
	&0 = -\bnabla\cdot \bi \jst(\bi x) \label{eq:def_pst}
	, \\
	&\bi \jst(\bi x) = \bi f(\bi x)\,\pst(\bi x) - \frac{1}{\beta}\bnabla \pst(\bi x) \label{eq:def_jst}
  . 
\end{eqnarray}
\endnumparts

%
%

Finding an explicit expression for $I[\rc]$ for Langevin processes may seem hopeless, but remarkably, for processes that satisfy detailed balance, $\jst(\bi x)\equiv0$,
Donsker and Varadhan showed quite some time ago  \cite{DonskerVaradhan1975a,DonskerVaradhan1975b,DonskerVaradhan1976,DonskerVaradhan1983} that the large deviation functional $I[\rc]$ is given by
\begin{eqnarray}
	I[\rc] 
  &= \frac{1}{\beta} \int \rmd \bm x \,
  \pst(\bm x) \left[
   \bnabla \sqrt{\frac{
       \rho(\bm x)
     }{
       \pst(\bm x)
     }
   }
  \right]^2 \\ 
  &= \frac{\beta}{4} \int \rmd \bi x\,\rho(\bm x) \left[
    \frac{1}{\beta}\bnabla\ln\frac{\rho(\bm x)}{\pst(\bm x)}
  \right]^2 \label{eq:LDF_DV}
  .
\end{eqnarray}
Note $I[\rho\!=\!\pst]=0$ and $I[\rc]\geq0$ for all $\rc$ as it should be. 

The Donsker-Varadhan result is limited to equilibrium situations. It is therefore natural to ask whether it can be generalized to non-equilibrium steady states (NESSs) where detailed balance is violated. NESSs come in a large variety since their stationary flux $\jst$ is, besides being divergenceless, largely arbitrary. We therefore expect that the large deviation functional $I[\rc]$ for a NESS will also depend on the stationary flux. We are hence motivated to introduce the empirical current
\begin{equation}\label{eq:def_mu}
 \bmu_T(\bi y;\bi x(\cdot)) = \frac{1}{T} \int_0^T \rmd t\, \dot \bi x(t) \,\delta(\bi y-\bi x(t))
\end{equation}
satisfying 
\begin{equation}\label{eq:limmu}
 \lim_{T \to \infty} \bmu_T(\bi y;\bi x(\cdot))=\bi \jst(\bi y)
\end{equation} 
in analogy to \eref{eq:limrho}. In further analogy to \eref{eq:contr_Ja}, the supplemented large deviation functional
\begin{equation}
  I[\rc,\mc] = -\lim_{T \to \infty}\frac{1}{T} \ln P_T[\rc,\mc]
\end{equation}
for the joint probability distribution $P_T[\rc,\mc]$ is suitable to derive all large deviation functions $J(b)$ of Brownian functionals of the type
\begin{eqnarray}
  b_T[\xc] &= \frac{1}{T}\int_0^T \rmd t\, \dot\bi x(t) \bi B(\bi x(t)) \nonumber\\
  &= \int \rmd\bi y\, \bi B(\bi y)\,\bmu_T(\bi y;\bi x(t)) \label{eq:B_T}
  ,
\end{eqnarray}
by contraction,
\begin{eqnarray}
  J(b) &= \underset{\mc\;|\;b=\int\!\rmd\bi y\,\bi B(\bi y)\,\bmu(\bi y)}{\min}\quad\underset{\rc}{\min}\quad I[\rc,\mc] \\
  &= \underset{\mc\;|\;b=\int\!\rmd\bi y\,\bi B(\bi y)\,\bmu(\bi y)}{\min}\quad I[\mc]\; \label{eq:contr_Jb}
  ,
\end{eqnarray}
where we did not write down the constraints $\bnabla\cdot\bmu(\bi y)=0$ and $\int\!\rmd\bi y\,\rho(\bi y)=1$ for clarity of notation. Since entropy productions are typically of the form \eref{eq:B_T}, the large deviation functional $I[\mc]$ is of particular interest.

%
%

In the present paper we elucidate the relation between the large deviation functional for the empirical density in stationary states of Langevin systems, the structure of the corresponding Fokker-Planck equation, and the entropy production in NESSs. In \sref{sec:det_bal} we first rederive the Donsker-Varadhan result \eref{eq:LDF_DV} using a path-integral formulation. Our derivation is a novel alternative to the mathematical treatment of Donsker and Varadhan \cite{DonskerVaradhan1975a,DonskerVaradhan1975b,DonskerVaradhan1976,DonskerVaradhan1983} and is meant to extend the accessibility to a less mathematically minded readership. In \sref{sec:ness} we obtain the large deviation functional $I[\rc, \mc]$ by generalizing the derivation to NESSs, i.e. to situations without detailed balance, and highlight the role of the probability current in a NESS. In \sref{sec:discussion} we show that $I[\rc, \mc]$ can be split into two contributions $\Ia[\mc]$ and $\Ina[\rc, \mc]$ linked to adiabatic and non-adiabatic entropy productions $\dSa$ and $\dSna$ \cite{Esposito2007,Esposito2010,Esposito2010a,VandenBroeck2010}. Moreover, we consider in some detail the contraction of $I[\rc, \mc]$ to $I[\rc]$ and $I[\mc]$ respectively and exemplarily discuss the contraction to $J(\Sa)$ where the adiabatic entropy production $\Sa$ is a typical instance of $b_T$ in \eref{eq:B_T}. Finally, in \sref{sec:example}, we illustrate our results by some examples.


%% file: detailed_balance.tex
\section{Systems with Detailed Balance}
\label{sec:det_bal}
For systems with detailed balance the external force derives from a potential, \mbox{$\bm f(\bm x) = -\bnabla V(\bm x)$}, and the stationary current is zero, $\jst(\bi x)\equiv0$. The stationary distribution is the equilibrium distribution $\pst(\bi x)=1/Z\exp(-\beta V(\bi x))$ with partition sum $Z$. We can hence write the force as
\begin{equation}
	\label{eq:f_from_st_detbal}
  \bm f(\bm x) = \frac{1}{\beta}\bnabla\ln\pst(\bm x)
\end{equation}
and may define the dynamics by fixing $\pst$ instead of $\bi f$.

To derive the Donsker-Varadhan result \eref{eq:LDF_DV} for the large deviation functional $I[\rc]$, we write $P_T[\rc]$ as the probability transformation of the probability density $P[\bm x(\cdot)]$ for observing a trajectory $\bm x(\cdot)$:
\begin{equation}
  \label{eq:P(x)}
    P_T[\rc]  =
    \int \mathcal{D} \bi x(\cdot) 
    P[\bi x(\cdot)]\,\delta[\rc - \rho_T(\,\cdot\,;\bi x(\cdot))].
\end{equation}
Using the integral representation of the $\delta$-functional, $P_T[\rc]$ may be written as 
\begin{equation}
\label{eq:CGF_motivation}
  \eqalign{
   P_T[\rc]  
   &= 
   \int \mathcal D q(\cdot)\, \exp\left[T \int \rmd y\, \rho(\bi y) q(\bi y)\right]
   Q_T[q(\cdot)],
 }
\end{equation}
with the cumulant generating function
\begin{equation}
  \label{eq:CGF1}
  \eqalign{
    Q_T[q(\cdot)] &= 
    \int \mathcal{D} \bi x(\cdot)\, 
    P_T[x(\cdot)] 
    \exp \left[
      -T \int \rmd \bi y\, \rho_T(\bi y;\,\bi x(\cdot))\, q(\bi y)
    \right] \\
    &= \int \mathcal{D} \bi x(\cdot)\, 
    P_T[x(\cdot)]
    \exp\left[ 
      -\int_0^T \rmd t\, q(\bi x(t))
    \right].
  }
\end{equation}
Like $P_T[\rc]$, for large $T$, $Q_T[q(\cdot)]$ can be written in a large deviation form: 
\begin{equation}
  \label{eq:lambda_def1}
  Q_T[q(\cdot)] = \exp\left[-T \lambda[q(\cdot)] + \mathrm o(T)\right]    
\end{equation}
with 
$
  \lambda[q(\cdot)] = - \lim_{T \to \infty} 1/T
   \ln   Q_T[q(\cdot)]
$.
Using this form of $Q_T[q(\cdot)]$ as well as \eref{eq:LDF_def} in \eref{eq:CGF_motivation} gives
\begin{equation}
  \label{eq:LF_trafo}
  \fl \exp\left[-T I[\rho(\cdot)] + \mathrm o(T)\right] 
   = \int \mathcal D q(\cdot) \exp \left\{
     -T \left[
       \lambda[q(\cdot)]  -\int \rmd \bi x\, \rho(\bi x) q(\bi x)
     \right] + \mathrm o(T)
   \right\}.
\end{equation}
For large $T$ the integral is dominated by its saddle-point and we find
\begin{equation}
  \label{eq:LF-trafo}
  I[\rc] = \min_{q(\cdot)} 
  \left(
    \lambda[q(\cdot)] -  \int \rmd\bi x\, \rho(\bi x) q(\bi x)
  \right).
\end{equation}
Hence, $-\lambda[-q(\cdot)]$ is the Legendre-Fenchel transform of $-I[-\rc]$
\cite{Touchette2009}.
Our strategy will be to determine $\lambda[q(\cdot)]$ first and than solve the minimization problem  \eref{eq:LF-trafo} to obtain $I[\rc]$.

The probability density of a trajectory   is given by 
\begin{equation}
  \label{eq:P(x)}
  P_T[\bi x(\cdot)] =  p(\bi x_0)   P[\bi x(\cdot)|\bi x_0],
\end{equation}
where  $p(\bi x_0)$ is the probability density of the initial point $\bi x_0 = \bi x(0)$ and $P[\bi x(\cdot)|\bi x_0]$ the conditional probability density of the trajectory $\{\bi x(\cdot) \}$,
\begin{equation}
	\fl
  P[\bi x(\cdot)|\bi x_0] = 
  \exp\left\{ 
    -\int_0^T \rmd t \left[
      \frac{\beta}{4} \left(
        \dot{\bi x}(t) + \bnabla V(\bi x(t))
      \right)^2
      - \frac{1}{2} \Delta V(\bi x(t))
    \right]
  \right\}.
\end{equation}
Expanding the square in the exponent, the mixed term 
\begin{equation}
 \int_0^T \rmd t\, \frac{\beta}{2}\, \dot{\bi x}(t) \bnabla V(\bi x(t)) 
    = \frac{\beta}{2} [V(\bm x(T)) - V(\bm x_0)]
\end{equation} 
gives rise to a boundary term that, for large $T$, contributes to the $\mathrm o(T)$ terms only. We may hence write  
\begin{equation}
  \label{eq:P(x|x0)}
  P[\bi x(\cdot)|\bi x_0] = 
  \exp\left\{ 
    -\int_0^T \rmd t \left[
      \frac{\beta}{4} 
        \dot{\bi x}^2(t) + U(\bi x(t))
    \right] + \mathrm o(T)
  \right\} 
\end{equation}
with $U(\bi x) =\beta / 4\, [\bnabla V(\bi x)]^2 - \frac{1}{2} \Delta V(\bi x)$. 

Inserting \eref{eq:P(x|x0)} into \eref{eq:CGF1} gives
\begin{equation}
  \label{eq:Q_pathintegral}
  Q_T[q(\bi y)] = \int \rmd \bi x_0 \, p(\bi x_0) \int \rmd \bi x_T\,
  G_q(\bm x_T,T |\bm x_0, 0)
\end{equation}
with 
\begin{equation}
  \label{eq:G_def}
  \fl
  G_q(\bm x_T, T |\bm  x_0, 0) = 
  \int_{(\bm x_0,0)}^{(\bm x_T,T)} \mathcal{D} \bi x(\cdot) 
  \exp \left[
    - \int_0^T \rmd t\,\left(
      \frac{\beta}{4} \dot{\bm x}^2(t) + U(\bm x(t))+ q(\bi x(t))
    \right)
  \right].
\end{equation}
Even for simple choices of $V(\bm x)$ this path-integral can not be solved for general $q(\cdot)$. Using the Feynman-Kac formula we may, however, obtain information on its large $T$ behavior  from the differential equation
\begin{equation}
  \label{eq:G_pde}
  \partial_t G_q(\bm x,t |\bm x_0, 0) = L_q G_q(\bm x,t |\bm x_0, 0)
\end{equation}
for the time evolution of $ G_q(\bm x,t |\bm  x_0, 0)$. The initial condition is $G_q(\bm x,0 |\bm  x_0, 0) = \delta(\bi x - \bi x_0)$ and $L_q$ is the so-called tilted generator of the Fokker-Planck dynamics
\begin{equation}
  \label{eq:L_q}
  L_q =\frac{1}{\beta} \Delta - U(\bi x) - q(\bi x).
\end{equation}

To determine $\lambda[q(\cdot)]$, we express the solution of \eref{eq:G_pde} in terms of the eigenvectors and eigenfunctions of $L_q$. We denote the right eigenvectors by $\phi^\nu_q$ and the corresponding eigenvalues by $\lambda^\nu_q$:
\begin{equation}
  L_q \phi^\nu_q = \lambda^\nu_q \phi^\nu_q.
\end{equation}
Owing to the symmetry of $L_q$, the left eigenvectors are simply given by ${\phi^\nu_q}^*$:
\begin{eqnarray} \label{eq:left_right_relation}
  L_q^+ {\phi^\nu_q}^* &=  \left[ \frac{1}{\beta} \Delta - U - q^*\right]  {\phi^\nu_q}^* \nonumber\\
  &= (L_q  \phi^\nu_q)^* = {\lambda^\nu_q}^* {\phi_q^\nu}^*.
\end{eqnarray}
Assuming that $L_q$ has a complete set of left and right eigenvectors\footnote{For a treatment without this assumption see \ref{sec:appendix_DE}.}, i.e.
\begin{equation}
  \label{eq:complete}
  \sum_\nu  \phi_q^\nu(\bi x) \phi_q^\nu(\bi y) = \delta(\bi x - \bi y),
\end{equation}
we can formally write the solution of the differential equation \eref{eq:G_pde} for arbitrary $q(\cdot)$ as
\begin{eqnarray}
  \label{eq:G_sol} \fl
  G_q(\bm x_T, T |\bm x_0, 0) &= \rme^{TL_q}\,\delta(\bi x_T - \bi x_0) \nonumber\\
  &= \rme^{TL_q} \sum_\nu  \phi_q^\nu(\bi x_T) \phi_q^\nu(\bi x_0) \nonumber\\
  &= \sum_\nu \rme^{T\lambda^\nu_q} \phi_q^\nu(\bi x_T) \phi_q^\nu(\bi x_0).
\end{eqnarray}
Inserting the above expression for $G_q(\bm x_T, T |\bm x_0, 0)$ into \eref{eq:Q_pathintegral} we get 
\begin{equation}
  \label{eq:Q_sol}
  Q_T[q(\bi y)] = \sum_\nu \rme^{T\lambda^\nu_q} 
  \int \rmd \bi x_0 \, p(\bi x_0)  \phi_q^\nu(\bi x_0)
  \int \rmd \bi x_T \phi_q^\nu(\bi x_T).
\end{equation}
Comparing with \eref{eq:lambda_def1}, we find for $T\to\infty$
\begin{equation}
 \lambda[q(\cdot)] = -\lambda_q^0,\label{eq:CGF}
\end{equation}
where $\lambda_q^0$ is the eigenvalue with the largest real part. For simplicity, we denote in the following the right eigenvector corresponding to this eigenvalue by $\phi_q$. The two integrals in \eref{eq:Q_sol} contribute to the $\mathrm o(T)$ term in \eref{eq:lambda_def1} only. Hence we do not need to know their value and can carry on without actually knowing the eigenvectors $\phi_q^\nu$.

Still, $\lambda[q(\cdot)]$ is hard to get for general $q(\cdot)$. However, we do not need an explicit expression. The Euler-Lagrange equation corresponding to the minimization problem \eref{eq:LF-trafo} is of the form 
\begin{equation}
 \left .\frac{\delta \lambda[q(\cdot)]}{\delta q(\bi x)}\right\vert_{q=\bar{q}}=\rho(\bi x)\; ,
\end{equation} 
where $\bar{q}(\cdot)$ denotes the $q(\cdot)$ that minimizes the r.h.s. of \eref{eq:LF-trafo}. 
Now, from standard first order perturbation theory 
\begin{eqnarray}
	\fl
  \lambda[q(\cdot) + \delta q(\cdot)] -  \lambda[q(\cdot)]
  &= - \int \rmd \bi x \, \phi_q(\bi x) 
  (L_{q + \delta q} - L_q) \phi_q(\bi x) + \Or(\|\delta q\|^2) \nonumber\\
  &= \int \rmd \bi x \, \phi_q(\bi x) 
  \delta q(\bm x) \phi_q(\bi x)
  + \Or(\|\delta q\|^2).
\end{eqnarray}
This leads to 
\begin{equation}
  \phi_{\bar{q}}^2(\bi x) = \rho(\bi x).
\end{equation}
 Since $\rho(\bi x)$ is positive, $\phi_{\bar q}$ is real and $L_{\bar q}$ has the same left and right eigenvectors. 

To finally determine $I[\rc]$, we plug the minimizing $\phi_{\bar q} = \sqrt \rho$ into  \eref{eq:LF-trafo} and find 
\begin{eqnarray}
  I[\rc] 
  &= \lambda[\bar q(\cdot)] - \int \rmd \bi x \, \bar q(\bm x) \rho(\bm x) \nonumber\\
  &= - \int \rmd \bi x\, \sqrt{ \rho(\bm x)} \left[
    L_{\bar q} + \bar q(\bm x)
  \right] \sqrt{\rho(\bm x)} \nonumber\\
  &= -\int \rmd \bi x\, \sqrt{ \rho(\bm x)} \left[
   \frac{1}{\beta}\Delta - U(\bm x)
  \right] \sqrt{\rho(\bm x)}.
\end{eqnarray}
Fortunately, the so far undetermined $\bar q(\cdot)$ drops out and we arrive at the result \eref{eq:LDF_DV} found by Donsker and Varadhan:
\begin{eqnarray}
  I[\rc] 
  &= -\int \rmd \bi x\, \sqrt{ \rho(\bm x)} \left[
    \frac{1}{\beta}\Delta 
    - \frac{\beta}{4}(\bnabla V(\bm x))^2
    + \frac{1}{2}\Delta V(\bm x)
  \right] \sqrt{\rho(\bm x)} \nonumber\\
  &= \int \rmd \bi x\,  \left[
    \frac{1}{\beta}\left(\bnabla \sqrt{\rho(\bm x)}\right)^2
    + \frac{\beta}{4}(\bnabla V(\bm x))^2 \rho(\bm x)
    + \frac{1}{2} \bnabla V(\bm x)\bnabla \rho(\bm x)
  \right] \nonumber\\
  &= \frac{\beta}{4} \int \rmd \bi x\,\rho(\bm x) \left(
    \frac{\bnabla \rho(\bm x)}{\beta\rho(\bm x)} + \bnabla V(\bm x)
  \right)^2 \nonumber\\
  &= \frac{\beta}{4} \int \rmd \bi x\,\rho(\bm x) \left[
    \frac{1}{\beta}\bnabla\ln\frac{\rho(\bm x)}{\pst(\bm x)}
  \right]^2
  .
\end{eqnarray}
Being normalized, $\rho(\bm x)$ tends to zero for $|\bm x|\to\infty$ and the boundary terms in the integration by parts do not contribute.

By substituting 
\begin{equation}
	\label{eq:def_jrho}
  \frac{\jrho(\bi x)}{\rho(\bi x)} = \frac{\jst(\bi x)}{\pst(\bi x)} - \frac{1}{\beta}\bnabla\ln\frac{\rho(\bi x)}{\pst(\bi x)}
\end{equation}
from the definitions \eref{eq:def_j} and \eref{eq:def_jst}, we find for $\jst(\bi x)\equiv0$ a compact form of the Donsker-Varadhan result,
\begin{equation}
	\label{eq:DV_jrho}
  I[\rc] = \frac{\beta}{4} \int \rmd \bm x \,\rho(\bm x)\,\left[\frac{\jrho(\bi x)}{\rho(\bi x)}\right]^2 
  .
\end{equation}



%% file: ness.tex
\section{Systems without Detailed Balance}
\label{sec:ness}
We now consider systems in which the force $\bi f(\bi x)$ does not derive from a potential. Consequently, it is $\jst(\bi x)\neq0$ and detailed balance is violated. Nevertheless, we can still relate the force to the stationary distribution $\pst$ and the stationary current $\jst$ by 
\begin{equation}
	\label{eq:f_from_st}
  \bi f(\bi x) = \frac{\jst(\bi x)}{\pst(\bi x)} + \frac{1}{\beta}\bnabla\ln\pst(\bi x)
  .
\end{equation}
As it is a hard problem to find $\pst$ and $\jst$ from \eref{eq:def_pst} and \eref{eq:def_jst} for a given force $\bi f$, it often is convenient to follow the reverse strategy and fix $\pst$ and $\jst$ and then determine the associated force $\bi f$ from the above equation. In this way, $\pst$ defines the conservative and $\jst$ the non-conservative part of $\bi f$.

In the following analysis we closely follow the lines in \sref{sec:det_bal}. The essential difference to the case with detailed balance becomes apparent by inspecting the probability density of a trajectory,
\begin{equation}
  \label{eq:P(x|x0)_2}
  P[\bi x(\cdot)|\bi x_0] = 
  \exp\left\{ 
    -\int_0^T \rmd t \left[
      \frac{\beta}{4} \left[
        \dot{\bi x}(t) - \bi f(\bi x(t))
      \right]^2
      + \frac{1}{2} \nabla \bi f(\bi x(t))
    \right]
  \right\}
  .
\end{equation}
The mixed term $\int_0^T \rmd t\, \dot \bi x \bi f(\bi x)$ is no longer negligible in the large $T$ limit but may be of order $T$. Therefore, the simplification applied in \eref{eq:P(x|x0)} can not be used here. This entails a more complicated form of the tilted generator which in turn implies that there is no such simple relation between the right and left eigenvectors as \eref{eq:left_right_relation}.

The fact that we now have $\jst$ in addition to $\pst$ to define the dynamics motivates to include the empirical current $\bmu$ defined in \eref{eq:def_mu} into our considerations and to investigate the form of the joint probability density $P_T[\rc, \mc]$. The corresponding large deviation form reads
 \begin{equation}
   \label{eq:LDF_ness}
   P_T[\rc, \mc] = \exp[-T I[\rc, \mc]
    + \mathrm o(T)].
 \end{equation}
The cumulant generating function of $P_T[\rc, \mc]$ now depends on two functions and is given by
\begin{equation}
  \label{eq:CGF}
  \fl
  \eqalign{
    Q_T[q(\cdot), \bi k(\cdot)] &= \int \mathcal{D} \bm x(\cdot)\,
    P_T[\bm x(\cdot)]
    \exp\left[ 
      - T \int \rmd \bm y\, q(\bm y) \rho(\bm y ; \xc )
       -T\int \rmd \bm y \, \bm k(\bm y)
       \bm \mu(\bm y ; \xc )
     \right]\\
     &=
  \int \mathcal{D} \bm x(\cdot)\, P_T[\bm x(\cdot)]
    \exp\left[
      - \int_0^T \rmd t\,[ q(\bm x(t)) + \dot{\bm x}(t) \bm k(\bm x(t))]
    \right].
  }
\end{equation}
The corresponding large deviation functional $\lambda[q(\cdot), \bi k(\cdot)]$ is defined by
\begin{equation}
  \label{eq:lambda_def_ness}
  Q_T[q(\cdot), \bi k(\cdot)] = \exp[-T \lambda[q(\cdot), \bi k(\cdot)] + \mathrm o(T)].
\end{equation}
At this point, it is convenient to substitute $q(\cdot)$ and $\bm k(\cdot)$ by two new functions  $\eta(\cdot)$ and $\bm \gamma(\cdot)$ such that
\begin{eqnarray}
  \label{eq:eta_gamma_def}
  q(\bm \gamma(\bi x),  \eta(\bi x)) &=
  -\frac{1}{2} \bnabla \bm \gamma(\bi x) - \frac{\beta}{2}\bi f(\bi x) \bm \gamma(\bi x)
  +\frac{\beta}{4} \bm \gamma^2(\bi x) + \eta(\bi x),\\
  \bm k( \bm \gamma(\bi x), \eta(\bi x)) &= \frac{\beta}{2} \bm \gamma(\bi x).
\end{eqnarray}
In analogy with \eref{eq:Q_pathintegral} we have
\begin{eqnarray}
\label{eq:Q_pathintegral_ness}
  Q_T[\bm \gamma(\cdot), \bi \eta(\cdot)] = 
  \int \rmd \bi x_0 \,p(\bi x_0)
  \int \rmd \bi x_T \,
  G_{\bm \gamma, \eta}(\bi x_T, T | \bm x_0, 0)
\end{eqnarray}
where now
\begin{eqnarray}   
	\fl
   G_{\bm \gamma, \eta}(\bi x_T, T |\bm x_0, 0) = 
   \int_{(x_0, 0)}^{(x_T, T)} \mathcal D \bi x(\cdot) 
   \exp \left\{
     - \int_0^T \rmd t\, \left[
       \frac{\beta}{4}[\dot \bi x(t) - \bi f(\bi x(t)) +\bm  \gamma(\bi x(t)) ]^2 
       \right. \right. \nonumber
       \\
       \left.\vphantom{\int} \left.
       +\frac{1}{2} \bnabla[\bi f(\bi x(t)) -\bm  \gamma(\bi x(t))] + \eta(\bi x(t))
     \right]
   \right\}.
\end{eqnarray}
Here, we have used \eref{eq:P(x|x0)_2} and \eref{eq:CGF}. The time evolution of $G_{\bm \gamma, \eta}(\bi x_T, T |\bm x_0, 0)$ is determined by
\begin{equation}
  \label{eq:pde_ness}
  \partial_t  G_{\bm \gamma, \eta}(\bi x, t | x_0, 0)
  = L_{\bm \gamma, \eta} G_{\bm \gamma, \eta}(\bi x, t | x_0, 0)
\end{equation}
with the tilted generator 
\begin{equation}
   L_{\bm \gamma, \eta} = \bnabla\left[ -\bi f + \bm \gamma + \frac{1}{\beta} \bnabla\right] - \eta.
\end{equation}
Again, the initial condition is $G_{\bm \gamma, \eta}(\bi x, 0 | \bm x_0, 0) = \delta(\bi x - \bi x_0)$.

We assume that $L_{\bm \gamma, \eta}$ has a complete set of right eigenvectors $\phi_{\bm \gamma, \eta}^\nu$ and left eigenvectors  $\psi^\nu_{\bm \gamma, \eta}$\footnote{For a treatment without this assumption see \ref{sec:appendix_DE}.} :
\begin{eqnarray}
  L_{\bm \gamma, \eta} \phi_{\bm \gamma, \eta}^\nu =   \lambda_{\bm \gamma, \eta}^\nu \phi_{\bm \gamma, \eta}^\nu, \\
  L^+_{\bm \gamma, \eta} \psi^\nu_{\bm \gamma, \eta} =   {\lambda_{\bm \gamma, \eta}^\nu}^* \psi_{\bm \gamma, \eta}^\nu ,
\end{eqnarray}
where the adjoint operator is given by 
\begin{equation}
 L_{\bm \gamma, \eta}^+ = [\bm f - \bm \gamma^* + \frac{1}{\beta}\bnabla]\bnabla - \eta^*\; .
\end{equation}  
The completeness of the eigenvector set is expressed by
\begin{eqnarray}
  \sum_\nu  {\psi_{\bm \gamma, \eta}^\nu}^*(\bi x) \phi_{\bm \gamma, \eta}^\nu(\bi y)
  = \delta(\bi x - \bi y).
\end{eqnarray}
As in the detailed balance case, we get as formal solution of \eref{eq:pde_ness}
\begin{eqnarray}
  \label{eq:Q_sol_ness}
  Q[\bm \gamma(\cdot), \eta(\cdot)] = 
  \sum_\nu \rme^{-\lambda_{\bm \gamma, \eta}^\nu T} 
  \int \rmd \bi x_0 \, p(\bi x_0) {\psi_{\bm \gamma, \eta}^\nu}^*(\bi x_0) 
  \int \rmd \bi x_T \, \phi_{\bm \gamma, \eta}^\nu(\bi x_T)
\end{eqnarray} 
and identify 
\begin{equation}
\lambda[\bm \gamma(\cdot), \eta(\cdot)] =  -\lambda_{\bm \gamma, \eta}^0,
\end{equation}
where $\lambda_{\bm \gamma, \eta}^0$ is the eigenvalue with the largest real part. For simplicity of notation the corresponding eigenvectors will be denoted by $\phi_{\bm \gamma, \eta}$ and $\psi_{\bm \gamma, \eta}$ in the following. 

To find $I[\rc,\mc]$ from $\lambda[\bm \gamma(\cdot), \eta(\cdot)]$ we have to solve the minimization problem 
\begin{eqnarray}
  \label{eq:max2}
  \fl
    I[\rc,\mc] &=
    \min_{q(\cdot), \bi k(\cdot)} \left(
      \lambda[q(\cdot), \bi k(\cdot)]
      - \int \rmd \bi x \, \rho(\bi x) q(\bi x)
      - \int \rmd \bi x \, \bmu(\bi x) \bi k(\bi x)
    \right) \nonumber\\
    &= \min_{\bm \gamma(\cdot), \bi \eta(\cdot)} \left(
      \lambda[\bm \gamma(\cdot), \eta(\cdot)]
      - \int \rmd \bi x \, \left\{
        \frac{\beta}{2}\bm\gamma(\bm x) 
        \left[
          \frac{\bnabla \rho(\bi x)}{\beta}
          + \bmu(\bm x)
          -\bm f(\bm x) \rho(\bm x)
        \right]\right. \right. \nonumber\\
        &\left.\left.\hspace{168pt}+\,\frac{\beta}{4}\bm \gamma^2(\bm x) \rho(\bm x)
        + \eta(\bm x) \rho(\bm x)
      \right\}
    \right).
\end{eqnarray}
Here, we integrated by parts to change $(\bnabla \bm \gamma) \rho$ into $- \bm \gamma \bnabla \rho$, where, due to the normalizability of $\rho(\bm x)$, the boundary terms do not contribute,.
 
Analogously to the minimization in $q(\cdot)$ for the detailed balance case, the minimization in $\eta(\cdot)$ leads in the present case to
\begin{equation}
  \psi^*_{\bar {\bm \gamma}, \bar \eta}(\bi x)  \phi_{\bar{\bm  \gamma}, \bar \eta}(\bi x)
  = \rho(\bi x).
\end{equation}
For the optimization in $\bm \gamma(\cdot)$ it is convenient to use the relation
\begin{equation}
  \lambda[\bm \gamma(\cdot), \eta(\cdot) ]= -\int \rmd \bm x \,
  [L^+_{\bm \gamma, \eta} \psi_{\bm \gamma, \eta}  (\bm x)]^*
  \phi_{\bm \gamma, \eta}(\bm x).
\end{equation}
With $L^+_{\bm \gamma + \delta \bm \gamma, \eta} -L^+_{\bm \gamma, \eta} = -\delta \bm \gamma^* \bnabla$ and making the substitution $\phi_{\bar{\bm \gamma}, \bar \eta} = \rho /  \psi_{\bar{\bm \gamma}, \bar \eta}^*$, the minimization in $\bm \gamma(\cdot)$ leads to
\begin{equation}
  \bnabla \psi^*_{\bar{\bm \gamma}, \bar \eta} = \psi^*_{\bar{\bm \gamma}, \bar \eta} \frac{\beta}{2}\,\bm A,
\end{equation}
where 
\begin{equation}
 \bm A := \frac{1}{\beta}\frac{\bnabla \rho}{\rho} + \frac{\bmu}{\rho} -\bm f + \bar{\bm \gamma}\; .
\end{equation} 

We are now able to put everything together and get from  \eref{eq:max2} 
\begin{eqnarray}
		\fl
    I[\rc, \mc] 
    &= \int \rmd \bm x \left\{
      - [L^+_{\bar{\bm \gamma}, \bar \eta} \, \psi_{\bar{\bm \gamma}, \bar \eta}]^* \, \phi_{\bar{\bm \gamma} , \bar \eta}
      - \frac{\beta}{2} \bar{\bm \gamma} \rho \bm A
      + \frac{\beta}{4}\rho \bar{\bm \gamma}^2
      - \bar{\eta} \rho
    \right\}\nonumber\\
    &=  \int \rmd \bm x \left\{
      \frac{\rho}{\psi^*_{\bar{\bm \gamma}, \bar \eta}}\left[
        \bm A - \frac{\bnabla \rho}{\beta \rho} 
        - \frac{\bmu}{\rho} - \frac{1}{\beta}\bnabla
      \right]\frac{\beta}{2}\bm A \psi^*_{\bar{\bm \gamma}, \bar \eta}
      - \frac{\beta}{2} \bar{\bm \gamma} \rho \bm A
      + \frac{\beta}{4}\rho \bar{\bm \gamma}^2
    \right\}\nonumber\\
    &=  \int \rmd \bm x \, \frac{\beta}{2}\left\{
      \rho \bm A^2 
      - \left[\frac{\bnabla \rho}{\beta} + \bmu \right] \bm A
      - \frac{\rho}{\beta} \bnabla \bm A - \frac{1}{2}\rho \bm A^2
      - \bar{\bm \gamma} \rho \bm A
      + \frac{1}{2} \rho \bar{\bm \gamma}^2
    \right\}\nonumber\\
    & = \frac{\beta}{4}  \int \rmd \bm x \, \rho \left[
      \frac{\bnabla \rho}{\beta \rho} + \frac{\bmu}{\rho} - \bm f
      \right]^2
      -  \int \rmd \bm x \, \bmu \bnabla \ln \psi^*_{\bar{\bm \gamma}, \bar \eta}.
\end{eqnarray}
Unlike the detailed balance case, the dependence on $\bar{\bm \gamma}$ and $\bar \eta$ does not vanish completely but survives in the last term involving  $\psi^*_{\bar{\bm \gamma}, \bar \eta}$. However, this term contributes to the $\mathrm o(T)$ terms in \eref{eq:lambda_def_ness} only. To prove this, we first integrate by parts:
\begin{equation}
\label{eq:last_term}
 \int \rmd \bm x \, \bmu(\bm x) \bnabla \ln \psi^*_{\bar{\bm \gamma}, \bar \eta}(\bm x)
 = -\int \rmd \bi x \, \ln \psi^*_{\bar{\bm \gamma}, \bar \eta}(\bi x) \bnabla \mu(\bi x).
\end{equation}
Next we recall the definition \eref{eq:def_mu} of $\bmu$ for an arbitrary trajectory $\{\bm y(\cdot) \}$ and find
\begin{equation}
  \label{eq:div_mu}
  \fl \bnabla \bmu(\bm x) 
  = \frac{1}{T} \int_0^T \rmd t \, \dot{ \bm y} \bnabla \delta(\bm x - \bm y(t))
  = \frac{1}{T}\left[
    \delta(\bm x - \bm y(0)) - \delta(\bm x - \bm y(T))
  \right].
\end{equation}
This in turn implies
\begin{equation}
  \label{eq:last_term_2}
   \int \rmd \bm x \, \bmu \bnabla \ln \psi^*_{\bar{\bm \gamma}, \bar \eta} 
   =  \frac{1}{T}\left[
    \ln \psi^*_{\bar{\bm \gamma}, \bar \eta}(\bm y(T)) -  \ln \psi^*_{\bar{\bm \gamma}, \bar \eta}(\bm y(0)) 
  \right]=\mathrm o(T).
\end{equation}
We are thus left with
\begin{equation}
	\label{eq:Isol_0}
  I[\rc, \mc] 
  = \frac{\beta}{4} \int \rmd \bm x \,
  \rho(\bm x)
  \left[
    \frac{1}{\beta} \bnabla \ln\rho(\bm x)
     + \frac{\bmu(\bm x)}{ \rho(\bm x)} 
     - \bm f (\bm x)
  \right]^2
  .
\end{equation}

$I[\rc, \mc]$ may be written in a number of alternative forms. Replacing $\bi f$ by using \eref{eq:f_from_st}, we find 
\begin{eqnarray}
	\fl
  I[\rc, \mc] &= \frac{\beta}{4} \int \rmd \bm x \,
  \rho(\bm x)
  \left[
    \left(\frac{\bmu(\bm x)}{\rho(\bm x)}-\frac{\jst(\bm x)}{\pst(\bm x)}\right) \;+\; \frac{1}{\beta}\bnabla\ln\frac{\rho(\bm x)}{\pst(\bm x)}
  \right]^2 \label{eq:Isol_mu}
  .
\end{eqnarray}
Upon repeated integrations by parts, the mixed term of the square can be shown to vanish,
\begin{eqnarray}
	\fl
  \Imix[\rc,\mc] &= \int \rmd \bm x \,\rho(\bm x)\,\left[\frac{\bmu(\bm x)}{\rho(\bm x)}-\frac{\jst(\bm x)}{\pst(\bm x)}\right] \cdot \bnabla\ln\frac{\rho(\bm x)}{\pst(\bm x)} \nonumber\\
  &= \oint \rmd \bm A \cdot \left[\left(\frac{\bmu}{\rho}-\frac{\jst}{\pst}\right)\rho\ln\frac{\rho}{\pst}\right] - \int \rmd \bm x \,\ln\frac{\rho}{\pst} \,\bnabla\cdot\left[\rho\,\left(\frac{\bmu}{\rho}-\frac{\jst}{\pst}\right)\right] \nonumber\\
  &= \oint \rmd \bm A \cdot \left[\left(\frac{\bmu}{\rho}-\frac{\jst}{\pst}\right)\rho\ln\frac{\rho}{\pst}\right] + \int \rmd \bm x \,\left[\jst\ln\frac{\rho}{\pst}\right]\cdot\bnabla\frac{\rho}{\pst} \nonumber\\
  &= \oint \rmd \bm A \cdot \left[\bmu\ln\frac{\rho}{\pst}\right] - \int \rmd \bm x \,\frac{\rho}{\pst}\,\bnabla\cdot\left[\jst\ln\frac{\rho}{\pst}\right] \nonumber\\
  &= \oint \rmd \bm A \cdot \left[\bmu\ln\frac{\rho}{\pst}\right] - \int \rmd \bm x \,\jst\cdot\bnabla\frac{\rho}{\pst} \nonumber\\
  &= \oint \rmd \bm A \cdot \left[\rho\left(\frac{\bmu}{\rho}\ln\frac{\rho}{\pst}-\frac{\jst}{\pst}\right)\right] = 0
	.
\end{eqnarray}
The integral over the boundary in the last line must vanishes due to the normalizability of the involved distributions. Thus, we are left with our final result
\begin{equation}
	\fl
	\label{eq:Isol_nu_final}
  I[\rc, \mc] = \frac{\beta}{4} \int \rmd \bm x \,
  \rho(\bm x)
  \left[
    \left(\frac{\bmu(\bm x)}{\rho(\bm x)}-\frac{\jst(\bm x)}{\pst(\bm x)}\right)^2 \;+\; \left(\frac{1}{\beta}\bnabla\ln\frac{\rho(\bm x)}{\pst(\bm x)}\right)^2
  \right]
  .
\end{equation}
As for the Donsker-Varadhan result, by substituting the current $\jrho$ from \eref{eq:def_jrho} into (\ref{eq:Isol_mu}), we obtain the compact form
\begin{eqnarray}
	\label{eq:I_vrho}
  I[\rc,\mc] &= \frac{\beta}{4} \int \rmd \bm x \,\rho(\bm x)\,\left[\frac{\bmu(\bi x)-\jrho(\bi x)}{\rho(\bi x)}\right]^2 
    .
\end{eqnarray}


%% file: discussion.tex
\section{Discussion}
\label{sec:discussion}

The large deviation functionals derived in the previous section characterize the distributions of empirical density and empirical current. They acquire a more intuitive meaning due to their relation to different forms of entropy production. In the first part of the  discussion we establish these connections. We then consider the question how to obtain the large deviation functionals $I[\rc]$ of the empirical density alone and $I[\bm \mu(\cdot)]$ of the empirical current alone for a non-equilibrium steady state. Finally, we discuss the large deviation function $J(\Sa)$ of the adiabatic entropy production $\Sa[\mc]$.

\subsection{Entropy production}
\label{sec:entropy_productions}

We start with the simpler case of systems obeying detailed balance. The stationary current is zero, $\jst(\bi x)\equiv0$,  and therefore, the dynamics is uniquely defined by the stationary distribution $\pst$ alone. If the system is initially out of equilibrium in a state $p\neq\pst$, it relaxes to equilibrium, and entropy is being produced with a rate \cite{Seifert2012}
\begin{equation}
	\label{eq:Sna_db}
  \dSi[\pc] = \beta\int \rmd \bm x \, p(\bi x) \left[\frac{\bi j_p(\bi x)}{p(\bi x)}\right]^2 
 .
\end{equation}
Here, $\bi j_p(\bi x)$ denotes the current corresponding to $p(\bi x)$ according to 
\begin{equation}
 \bi j_p(\bi x)=-\frac{1}{\beta} \, p(\bi x)\bnabla \ln\frac{p(\bi x)}{\pst(\bi x)}\, . 
\end{equation} 
Comparison with the Donsker-Varadhan result \eref{eq:DV_jrho} reveals \cite{Maes2008,Wynants2010}
\begin{equation}\label{eq:ISna_detbal}
 I[\rho(\cdot)] = \frac{1}{4} \dSi[\rho(\cdot)],
\end{equation} 
i.e., the large deviation functional $I[\rc]$ is, up to a constant factor, nothing but the entropy production of the system when being in the state $\rc$ instead of the equilibrium state $\pst$. For large $T$, the entropy production hence quantifies how unlikely a deviation of the empirical density from the true equilibrium distribution is. 

For systems without detailed balance it has been shown \cite{Esposito2010,Esposito2010a,VandenBroeck2010,Esposito2007} that the entropy production rate $\dSi$ may be subdivided into two contributions, the so-called adiabatic and non-adiabatic parts, $\dSi=\dSa+\dSna$, where 
\begin{eqnarray}
 \label{eq:Sa}
  \dSa[\pc] &= \beta\int \rmd \bm x \, p(\bi x) \left[\frac{\bi \jst(\bi x)}{\pst(\bi x)}\right]^2,\\
\label{eq:Sna}
  \dSna[\pc] &= \beta\int \rmd \bm x \, p(\bi x) \left[\frac{\bi j_p(\bi x)}{p(\bi x)} - \frac{\bi \jst(\bi x)}{\pst(\bi x)}\right]^2\; .
\end{eqnarray} 
The non-adiabatic part $\dSna$ is the generalization of \eref{eq:Sna_db} and describes the entropic cost to relax to the stationary state. Correspondingly, it vanishes for $\bi j_p(\bi x)=\bi \jst(\bi x)$ and $p(\bi x)=\pst(\bi x)$. The adiabatic part $\dSa$, on the other hand, remains non-zero even in the steady state and characterizes the dissipation necessary to maintain stationarity away from equilibrium. The two contributions therefore address the two basic mechanisms that commonly go along with non-equilibrium situations: driving and out of equilibrium boundary conditions respectively \cite{Esposito2010,Esposito2010a,VandenBroeck2010,Esposito2007}.

Using \eref{eq:def_jrho} we may write the large deviation functional \eref{eq:I_vrho} in the form 
\begin{equation}
 \fl
  I[\rc, \mc] = \frac{\beta}{4} \int \rmd \bm x \,
  \rho(\bm x)
  \left[
    \left(\frac{\bmu(\bm x)}{\rho(\bm x)}-\frac{\jst(\bm x)}{\pst(\bm x)}\right)^2 \;+\; \left(\frac{\jrho(\bi y)}{\rho(\bi y)} - \frac{\jst(\bi y)}{\pst(\bi y)}\right)^2
  \right]
  ,
\end{equation}
which suggests to split it into the two contributions
\numparts
\begin{eqnarray}
  \Ia[\rc,\mc] &= \frac{\beta}{4} \int \rmd \bm x \, \rho(\bi x) \left[ \frac{\bmu(\bm x)}{\rho(\bm x)}-\frac{\jst(\bm x)}{\pst(\bm x)} \right]^2 \label{eq:Ia} ,\\
	\Ina[\rc] &= \frac{\beta}{4} \int \rmd \bm x \, \rho(\bi x) \left[ \frac{\jrho(\bi y)}{\rho(\bi y)} - \frac{\jst(\bi y)}{\pst(\bi y)} \right]^2 \label{eq:Ina} , 
\end{eqnarray}
\endnumparts
such that
\begin{equation}
  I[\rc,\mc] =  \Ia[\rc,\mc] + \Ina[\rc]
  .
\end{equation}
Comparing \eref{eq:Ina} and \eref{eq:Sna} we find in analogy to \eref{eq:ISna_detbal}
\begin{equation}
 \Ina[\rc] = \frac{1}{4} \dSna[\rc]\, .
\end{equation} 
The second term in $I[\rc,\mc]$, $\Ina[\rc]$, is hence (up to a constant factor) equal to the non-adiabatic part of the entropy production that accounts for deviations of $\rho$ from $\pst$. Accordingly, it characterizes the difference between empirical and true stationary distribution and is independent of $\bmu$.

The correspondence between $\Ia[\rc,\mc]$ and $\dSa[\pc]$ as defined in \eref{eq:Sa} is not quite as close. The reason is twofold. Firstly, $\dSa[\pc]$ is a functional of $p$ alone and has no dependence on a current. Secondly, $\Ia[\rc,\mc]$ must be zero for $\rho=\pst$ and $\bmu=\jst$, whereas $\dSa[\pc]$ has to remain non-zero even in the steady state. To connect $\dSa[\pc]$ with $\Ia[\rc,\mc]$, we have hence (in addition to the prefactor $1/4$) to make the replacement $\jst/\pst \to \jst/\pst-\bmu/\rho$. 
The first contribution to $I[\rc,\mc]$, $\Ia[\rc,\mc]$, hence measures the distance between the empirical and the true stationary current. Evaluated at $\bmu=0$ it gives (up to a constant factor) the adiabatic entropy production in state $\rho$.

We note in passing that the above expressions may be formally simplified by introducing the local mean velocity $\bi v=\bi j/p$, see, e.g., \cite{Seifert2012}, giving rise to $\bnu=\bmu/\rho$, $\vrho=\jrho/\rho$, $\vst=\jst/\pst$, and the large deviation functionals $I[\rc,\nc]$, $\Ia[\rc,\nc]$, and $\Ina[\rc]$. We then find , e.g., from $\Ia[\rc,\nc]=0$ immediately $\bnu=\vst$, completely independent of $\rho$. 


\subsection{Contractions}
\label{sec:marginalization}
The large deviation functional $I[\rc, \mc]$ characterizes the joint probability distribution $P[\rc,\mc]$ for 
empirical density and empirical current. In many situations one is interested in the distribution of either the density or the current alone. These distributions are obtained by marginalization
\begin{equation}
 P[\rc]=\int \mathcal{D}\mc P[\rc,\mc],\quad P[\mc]=\int \mathcal{D}\rc P[\rc,\mc] \; .
\end{equation} 
If the involved probability distributions obey large deviation principles, marginalization naturally transforms into contraction \cite{Touchette2009}, i.e.
\begin{equation}
 I[\rc]= \underset{\bmu}{\min}\; I[\rc, \mc], \qquad I[\mc]= \underset{\rho}{\min}\; I[\rc, \mc]\; .
\end{equation} 
We start with the determination of $I[\rc]$. To find $I[\rc]=I[\rc,\bmu\!=\!\bbmu(\rho)]$, we have to determine the \emph{optimal} current $\bbmu(\rho)$ that minimizes $I[\rc, \mc]$ for every empirical distribution $\rho$. Since $I[\rc,\mc]$ depends on $\bmu$ solely via $\Ia[\rc,\mc]$, we need to minimize $\Ia$ only. 
In view of \eref{eq:last_term}, (\ref{eq:div_mu}) and \eref{eq:last_term_2}, this minimization has to be done under the constraint $\bnabla\cdot\bmu=0$: 
\begin{equation}
  I[\rc] = \underset{\bmu|\bnabla\cdot\bmu=0}{\min} I[\rc, \mc] \; .
\end{equation}
Including the constraint with a Lagrange multiplier field $\kappa(\bi x)$ yields the Euler-Lagrange equation
\begin{equation}
  \rho(\bi x)\left(\frac{\bbmu(\bi x)}{\rho(\bi x)}-\frac{\jst(\bi x)}{\pst(\bi x)}\right)\frac{1}{\rho(\bi x)} - \bnabla\kappa(\bi x) = 0
  .
\end{equation}
The optimal current can hence be determined from the equations
\numparts
\begin{eqnarray}
  \bbmu(\bi x;\rho) &= \frac{\rho(\bi x)}{\pst(\bi x)}\,\jst(\bi x) + \rho(\bi x)\bnabla\kappa(\bi x) \label{eq:mu_opt}
  ,\\
  \Delta\kappa(\bi x) &= -\frac{\jst(\bi x)}{\pst(\bi x)}\cdot\bnabla\ln\frac{\rho(\bi x)}{\pst(\bi x)} - \bnabla\kappa(\bi x)\cdot\bnabla\ln\rho(\bi x) \label{eq:kappa_opt}
  ,
\end{eqnarray}
\endnumparts
where the second line follows from $\nabla\cdot\bbmu=0$. The resulting large deviation functional reads
\begin{equation}\label{eq:res_cont}
  I[\rc] = \Ina[\rc] + \frac{\beta}{4}\int \rmd \bi x \, \rho(\bi x)\left[\bnabla\kappa(\bi x)\right]^2
  .
\end{equation}

For systems with detailed balance we have $\jst(\bi x)\equiv0$, and $\kappa\equiv 0$ solves \eref{eq:kappa_opt}. From \eref{eq:res_cont} we then recover the Donsker-Varadhan result in the form \eref{eq:DV_jrho}. Remarkably, in this case the optimal current is zero for all $\rho$. For $\jst(\bi x)\neq0$ the determination of $I[\rc]$ remains a challenging problem.

We continue with the determination of $I[\mc]$. Now we need the optimal density $\bar\rho$ that minimizes $I[\rc, \mc]$:
\begin{equation} \label{eq:contr_Imu}
  I[\mc] = \underset{\rho\,|\int\!\rho\,\rmd\bi x\,=\,1}{\min} I[\rc, \mc] = I[\bar\rho(\bmu),\mc]
  .
\end{equation}
The Euler-Lagrange equation for $\bar\rho$ reads
\begin{equation}
  \left(\frac{\bmu}{\bar\rho}\right)^2 - \left(\frac{\jst}{\pst}\right)^2 + \frac{1}{\beta^2}\left(\left(\bnabla\ln\bar\rho\right)^2 - \left(\bnabla\ln\pst\right)^2 + 2\Delta\ln\frac{\bar\rho}{\pst}\right) = \kappa \label{eq:ele_brho}
  ,
\end{equation}
where $\kappa$ is the Lagrange parameter to be determined from the constraint \mbox{$\int\!\bar\rho(\bi x;\kappa)\,\rmd\bi x=1$.}
Solving the above Euler-Lagrange equation analytically is barely possible and no progress seems possible.

However, as discussed below \eref{eq:B_T}, we stress that $I[\mc]$ is particularly relevant for contractions to large deviation functions of entropy productions. Consider for example the adiabatic entropy production $\Sa[\xc]$ on the trajectory level \cite{Seifert2012,SpeckSeifert05JoPAMaG}
\begin{equation}
  \Sa[\xc] = \beta\int_{0}^{T}\rmd t \; \dot\bi x(t)\cdot\frac{\jst(\bi x(t))}{\pst(\bi x(t))} \label{eq:Sa_x}
  .
\end{equation}
Note that as opposed to the spatial average $\dSa$ in \eref{eq:Sa}, $\Sa[\xc]$ is a fluctuating quantity and satisfies a fluctuation theorem, see \cite{Seifert2012,Speck2012} for details. Identifying \mbox{$B(\bi y)=\jst(\bi y)/\pst(\bi y)$} in \eref{eq:B_T}, we can rewrite $\Sa[\xc]$ as
\begin{eqnarray}
  \Sa[\mc] &= \beta T\int\rmd\bi y\,\bmu(\bi y) \cdot \frac{\jst(\bi y)}{\pst(\bi y)} \label{eq:Sa_mu}
  .
\end{eqnarray}

We now use $\Sa[\mc]$ as an example to illustrate the contraction to large deviation functions. Having performed the contraction in \eref{eq:contr_Imu} to obtain $I[\mc]$, the subsequent contraction 
\begin{equation}
  J(\Sa) = \underset{\bmu\,|\,\bnabla\cdot\bmu=0\,,\;\Sa=\Sa[\bmu]}{\min} I[\mc]  \label{eq:contr_JSa}
\end{equation}
would yield the large deviation function $J(\Sa)$ for the adiabatic entropy production. But, due to the non-linearity of the Euler-Lagrange equations \eref{eq:ele_brho} for the contraction to $I[\mc]$, analytical progress in finding $J(\Sa)$ will barely be possible. Solving the Euler-Lagrange equations \eref{eq:ele_brho} numerically, however, may be possible, but the resultant parametrical dependency of $\bar\rho$ and $\kappa$ on $\bmu$ hinders the formulation of the Euler-Lagrange equation for the above contraction and complicates the determination of $J(\Sa)$ even in a numerical treatment.

An alternative strategy, where some progress can be made, is to flip the order of contractions and first perform
\begin{equation}
  \Ia[\rc,\Sa] = \underset{\bmu\,|\,\bnabla\cdot\bmu=0\,,\;\Sa=\Sa[\bmu]}{\min} \Ia[\rho,\mc] \label{eq:contr_ISa}
\end{equation}
and then substitute the optimal density $\bar\rho$.
The Euler-Lagrange equation of the above contraction reads
\begin{equation}
  \left(\frac{\bmu(\bi x)}{\rho(\bi x)} - \frac{\jst(\bi x)}{\pst(\bi x)}\right) - \bnabla\kappa_1(\bi x) + \kappa_2\frac{\jst(\bi x)}{\pst(\bi x)} = 0 \label{eq:ele_bbmu_Sa}
  ,
\end{equation}
where $\kappa_1(\bi x)$ is the Lagrange function for the constraint $\bnabla\cdot\bmu=0$ and $\kappa_2$ the Lagrange parameter for the constraint $\Sa=\Sa[\bmu]$ from \eref{eq:contr_JSa}. The optimal current hence reads
\begin{equation}
  \bbmu(\bi x) = \left(1-\kappa_2\right)\frac{\rho(\bi x)}{\pst(\bi x)}\jst(\bi x) + \rho(\bi x)\bnabla\kappa_1(\bi x) \label{eq:bbmu_Sa}
\end{equation}
which substituted into $\Ia[\rc,\mc]$ yields
\begin{equation}
  \Ia[\rc,\Sa] = \frac{\beta}{4} \int \rmd\bi x\, \rho(\bi x)\left[\bnabla\kappa_1(\bi x) - \kappa_2\frac{\jst(\bi x)}{\pst(\bi x)}\right]^2 \label{eq:Ia_rhoSa}
  .
\end{equation}
The Lagrange function $\kappa_1$ and the Lagrange parameter $\kappa_2$ has to be determined from the equations for the respective constraints,
\numparts
\begin{eqnarray}
  0 &= \Delta\kappa_1(\bi x) + \bnabla\kappa_1(\bi x)\cdot\bnabla\ln\rho(\bi x) + (1-\kappa_2)\,\frac{\jst(\bi x)}{\pst(\bi x)}\cdot\bnabla\ln\frac{\rho(\bi x)}{\pst(\bi x)}, \label{eq:kappa1} \\
  \Sa &= T\beta\int\rmd\bi y\,\rho(\bi y)\,\frac{\jst(\bi y)}{\pst(\bi y)}\cdot\left[(1-\kappa_2)\frac{\jst(\bi y)}{\pst(\bi y)} + \bnabla\kappa_1(\bi y)\right]. \label{eq:kappa2}
\end{eqnarray}
\endnumparts
Substituting the empirical adiabatic entropy production rate $\dSar$ from \eref{eq:Sa}, we can rewrite \eref{eq:kappa2} as
\begin{equation}
	\fl
  \frac{\Sa}{T} = \dSar - \frac{\beta}{\kappa_2}\int \rmd\bi y\, \rho\left[\bnabla\kappa_1 - \kappa_2\frac{\jst}{\pst}\right]^2 + \frac{T\beta}{\kappa_2}\int\rmd\bi y\,\rho\,\left[\left(\bnabla\kappa_1\right)^2-\kappa_2\bnabla\kappa_1\cdot\frac{\jst}{\pst}\right]
\end{equation}
and find from substituting \eref{eq:bbmu_Sa} and \eref{eq:Ia_rhoSa}
\begin{eqnarray}
  \Ia[\rc,\Sa] &= \frac{\kappa_2}{4}\left(\dSar - \frac{\Sa}{T}\right) + \frac{\beta}{4}\int\rmd\bi y\,\rho\,\bnabla\kappa_1\cdot\left(\frac{\bbmu}{\rho}-\frac{\jst}{\pst}\right) \nonumber\\
  &= \frac{\kappa_2}{4}\left(\dSar - \frac{\Sa}{T}\right) - \frac{\beta}{4}\int\rmd\bi y\,\frac{\rho}{\pst}\,\jst\cdot\bnabla\kappa_1 \label{eq:Ia_rho_Sa}
  ,
\end{eqnarray}
where the second line follows after two integrations by parts. The remaining quantities to be determined are $\kappa_1(\bi x;\Sa)$ and $\kappa_2(\Sa)$ from \eref{eq:kappa1} and \eref{eq:kappa2}, and $\bar\rho(\kappa)$ from \eref{eq:ele_brho} where $\bmu=\bbmu(\Sa)$ needs to be substituted from \eref{eq:bbmu_Sa} and $\kappa$ follows from the constraint \mbox{$\int\!\bar\rho(\bi x;\kappa)\,\rmd\bi x=1$.}  Hence, by first contracting to $I[\rc,\Sa]$ and then to $J(\Sa)$, as opposed to first contract to $I[\mc]$ and then to $J(\Sa)$, all necessary equations to determine $J(\Sa)$ are known and a numerical treatment is possible.

%% file: example.tex
\section{Illustration}
\label{sec:example}
To illustrate our findings, we first consider a simple example without detailed balance given by the force
\begin{equation}
	\label{eq:ex1_f}
  \bi f(\bi r) = -ar\bi e_r + br\bi e_\varphi
\end{equation}
with the constants $a>0$ and $b$. The stationary distribution and the stationary current are
\numparts
\begin{eqnarray}
	\label{eq:ex1_pst}
  \pst(r) &= \frac{\beta a}{2\pi}\,\exp\left[-\beta\,\Phi(r)\right] \,,\quad &\Phi(r) = \frac{a}{2}\,r^2 \\
	\label{eq:ex1_jst}
  \jst(r) &= \pst(r)\,\vst(r) \,,\quad &\vst(r) = br\,\bi e_\varphi
\end{eqnarray}
\endnumparts
and we can thus also write
\begin{eqnarray}
 \bi f(r) &= \frac{1}{\beta}\frac{\partial}{\partial r}\ln\pst(r)\,\bi e_r + \frac{\jst(r)}{\pst(r)} \label{eq:ex1_f}\\
 &= -\partial_r\Phi(r)\,\bi e_r + \vst(r) .
\end{eqnarray}
The conservative force $-\partial_r\Phi(r)$ points to the origin and keeps the dynamics bounded, the non-conservative force $\vst(r)$ generates a circular current and violates the detailed balance condition.

For a numerical illustration, we fix the parameters to be $a=1$, $b=2$, $T=500$, $\beta=2$ and numerically solve the Langevin equation \eref{eq:langevin1} with $\bi f(r)$ from \eref{eq:ex1_f} using $10^5$ timesteps to obtain an ensemble of $10^3$ trajectories $\{\bi r(\cdot)\}$. The initial values $\bi r(0)$ are drawn from the stationary distribution \eref{eq:ex1_pst} in order to save the relaxation time and have the system in the NESS the whole time. To each trajectory, we determine the empirical density $\rc$ and current $\mc$ from the definitions \eref{eq:def_rho} and \eref{eq:def_mu} and calculate from that the values of $\Ia[\rc,\mc]$ and $\Ina[\rc]$ using \eref{eq:Ia} and \eref{eq:Ina}.

\begin{figure}
	\centering
  \includegraphics[width=0.49\textwidth]{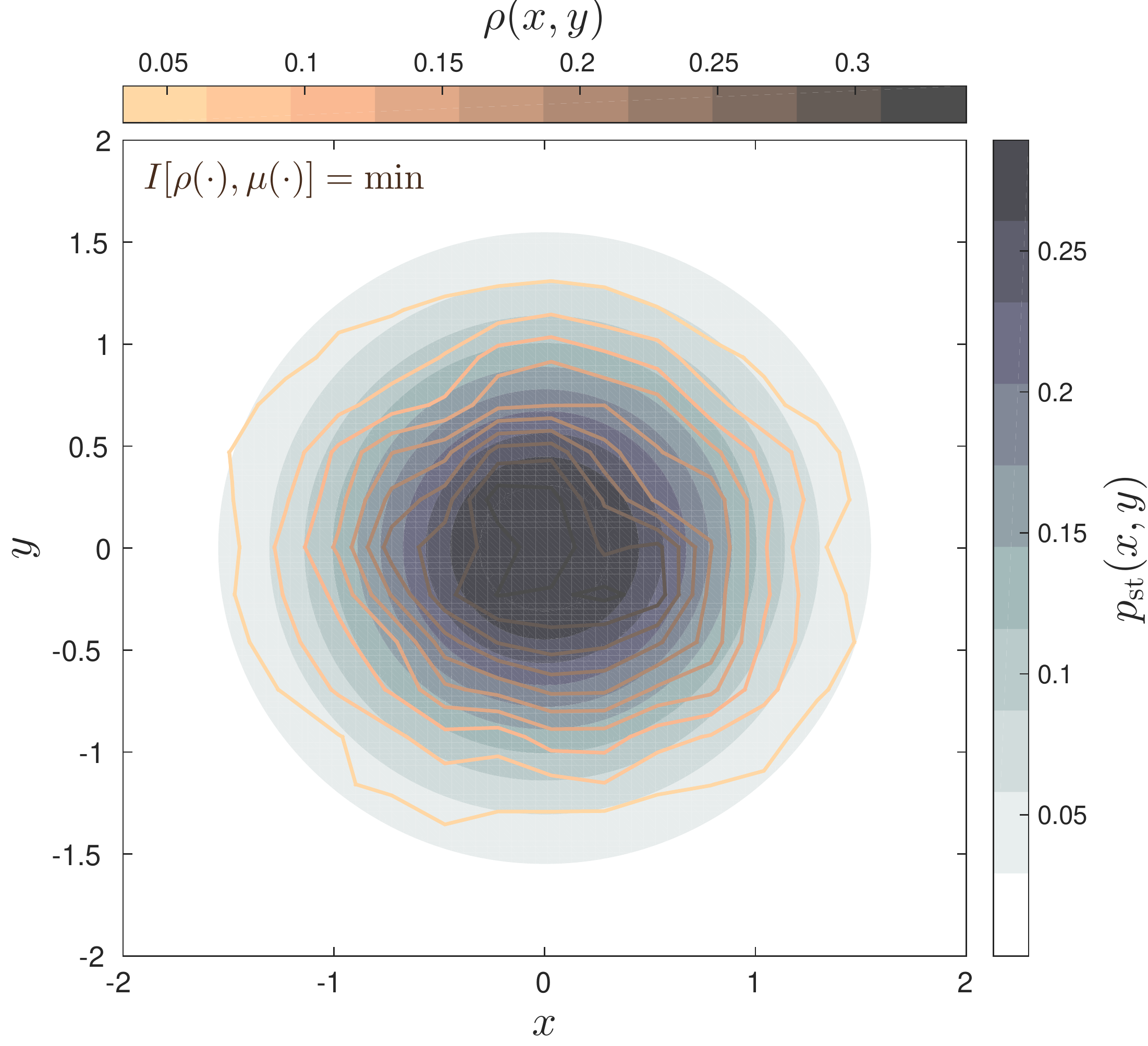}
  \includegraphics[width=0.49\textwidth]{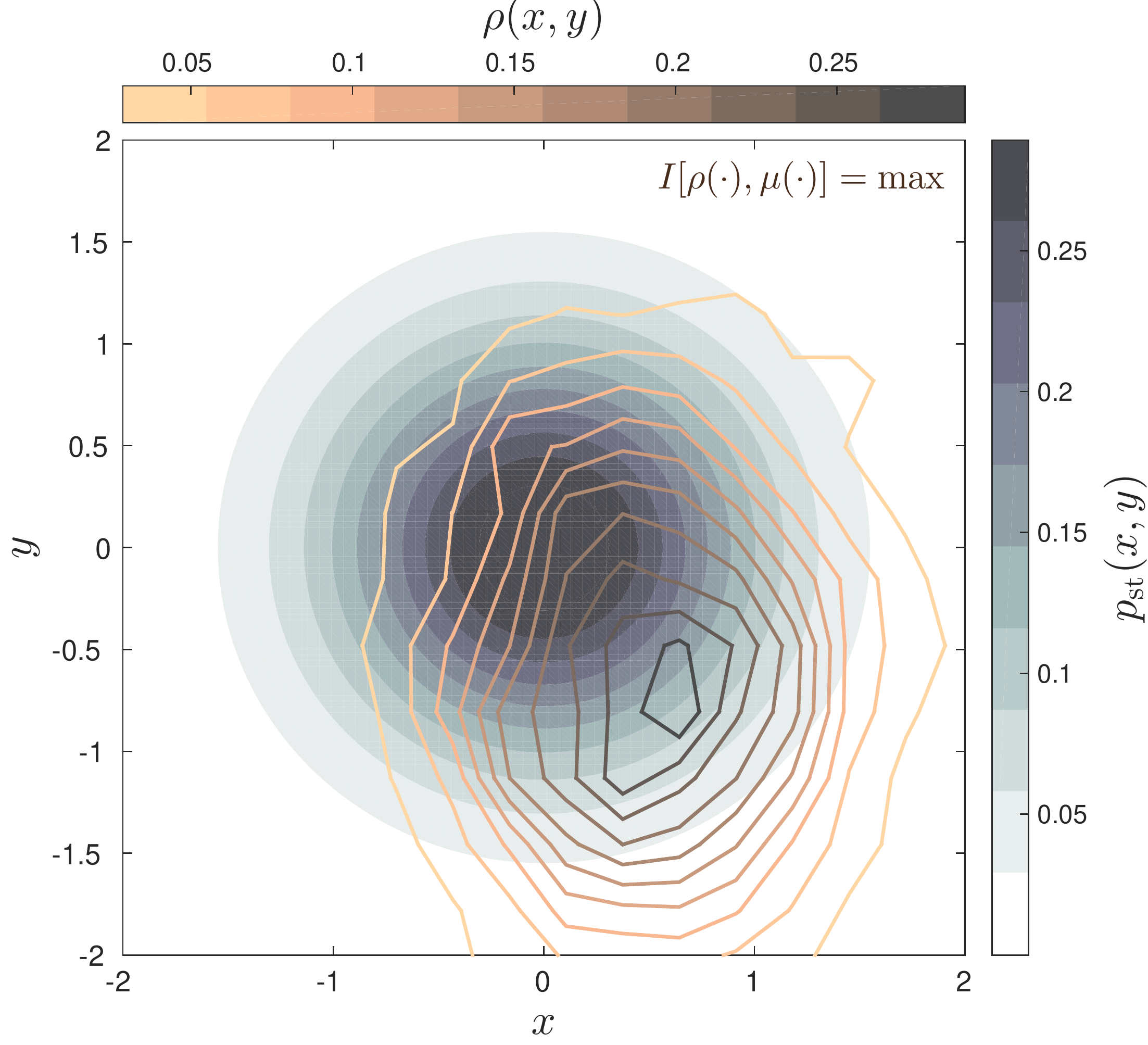}
  \caption{Two realisations of the empirical density $\rho$ that lead to particular low (left) and high (right) values for $I[\rc,\mc)]$. The contours comply with the value of $\rho(x,y)$, the shading represents $\pst(x,y)$. The parameters are $a=1$, $b=2$, $T=500$ and $\beta=2$.}
	\label{fig:rho_sample_minmax}
\end{figure}

\begin{figure}
	\centering
  \includegraphics[width=0.49\textwidth]{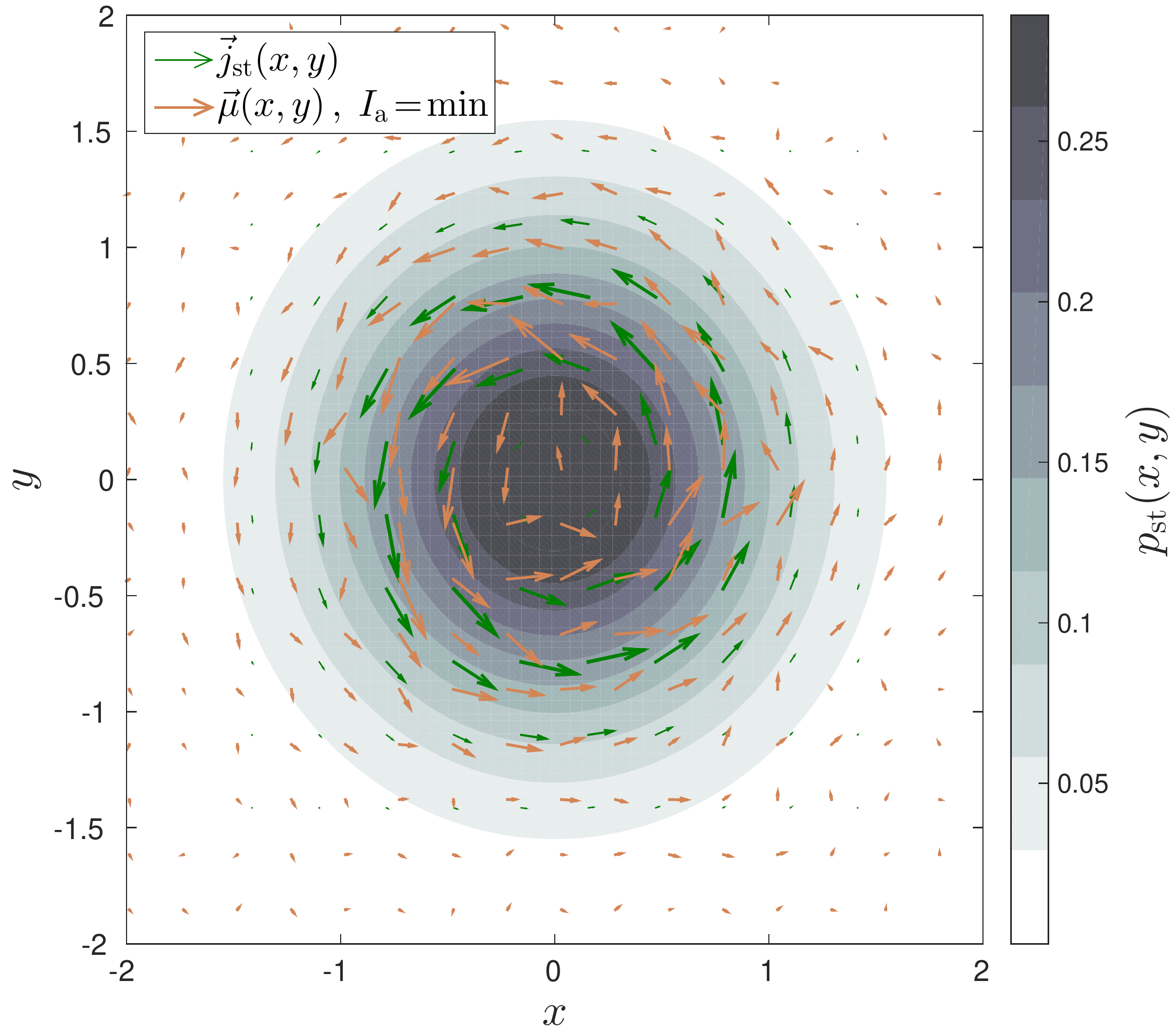}
  \includegraphics[width=0.49\textwidth]{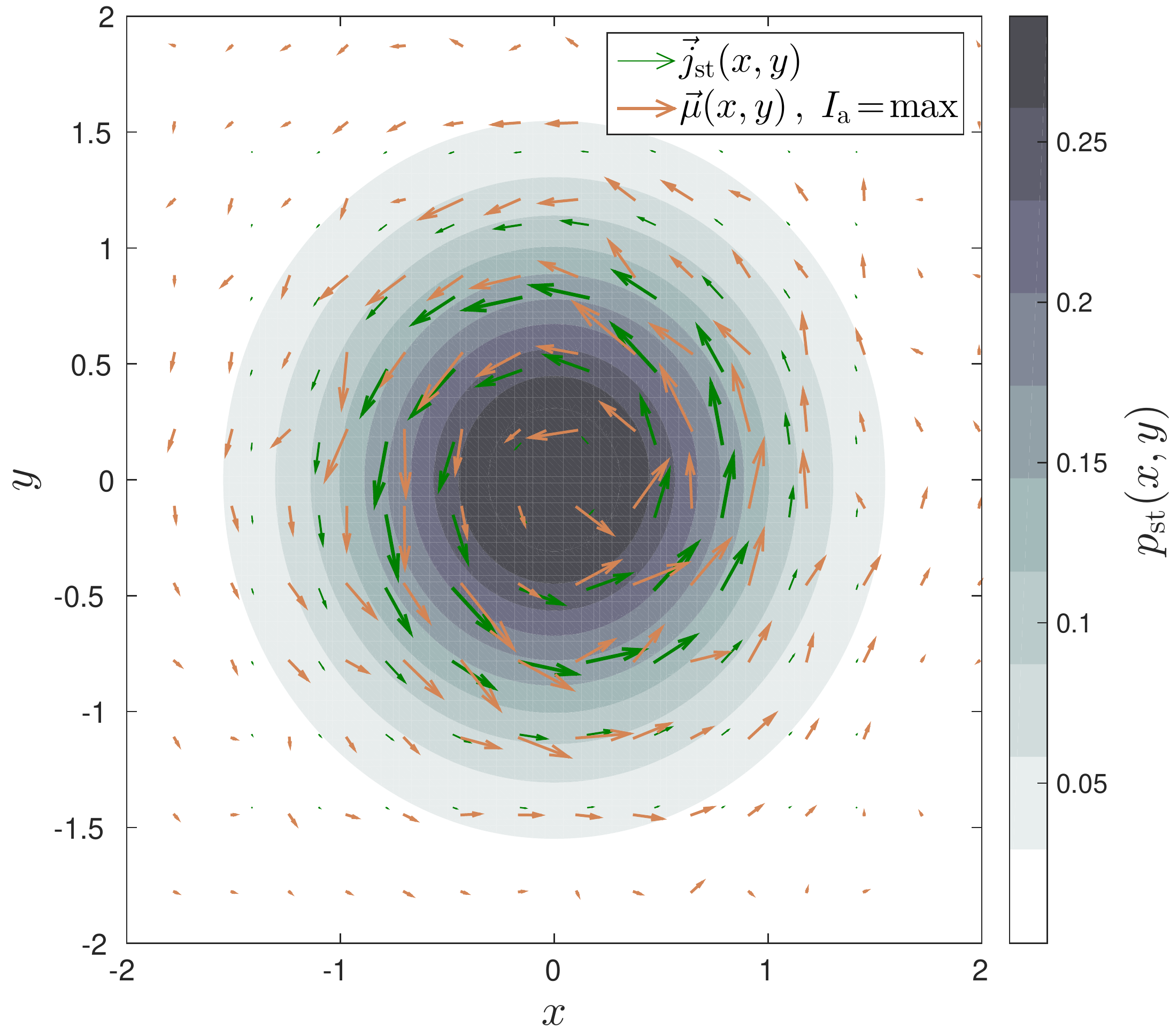}
  \caption{Two realisations of the empirical density $\bmu$ that lead to particular low (left) and high (right) values for $I[\rc,\mc]$. The arrows point in the direction of $\bmu(x,y)$ resp. $\jst(x,y)$, the arrow length complies with $|\bmu(x,y)|$ resp. $|\jst(x,y)|$. The parameters are $a=1$, $b=2$, $T=500$ and $\beta=2$.}
	\label{fig:mu_sample_minmax}
\end{figure}

In \fref{fig:rho_sample_minmax} we show a typical and a rare realization of $\rho$, and in \fref{fig:mu_sample_minmax} a typical and a rare realization of $\bmu$. The {\it typical} realizations are selected by picking from the ensemble the $\rho$ and $\bmu$ with the minimum value of $I[\rc,\mc]$ respectively, and correspondingly, the {\it rare} realizations are qualified by the maximum value of $I[\rc,\mc]$ within the ensemble. Note that in order to select the typical and rare realization of $\bmu$, we only need $\Ia[\rc,\mc]$. We see that the typical realization of $\rho$ is close to $\pst$, whereas the rare realization clearly deviates from $\pst$. In contrast, both realizations of $\bmu$ are close to $\jst$, implying that $I[\rc,\mc]$ is much sharper with respect to $\bmu$ than to $\rho$, that is, the system predominantly fluctuates with regard to position and barely with regard to direction of movement.

\begin{figure}
	\centering
  \includegraphics[width=0.49\textwidth]{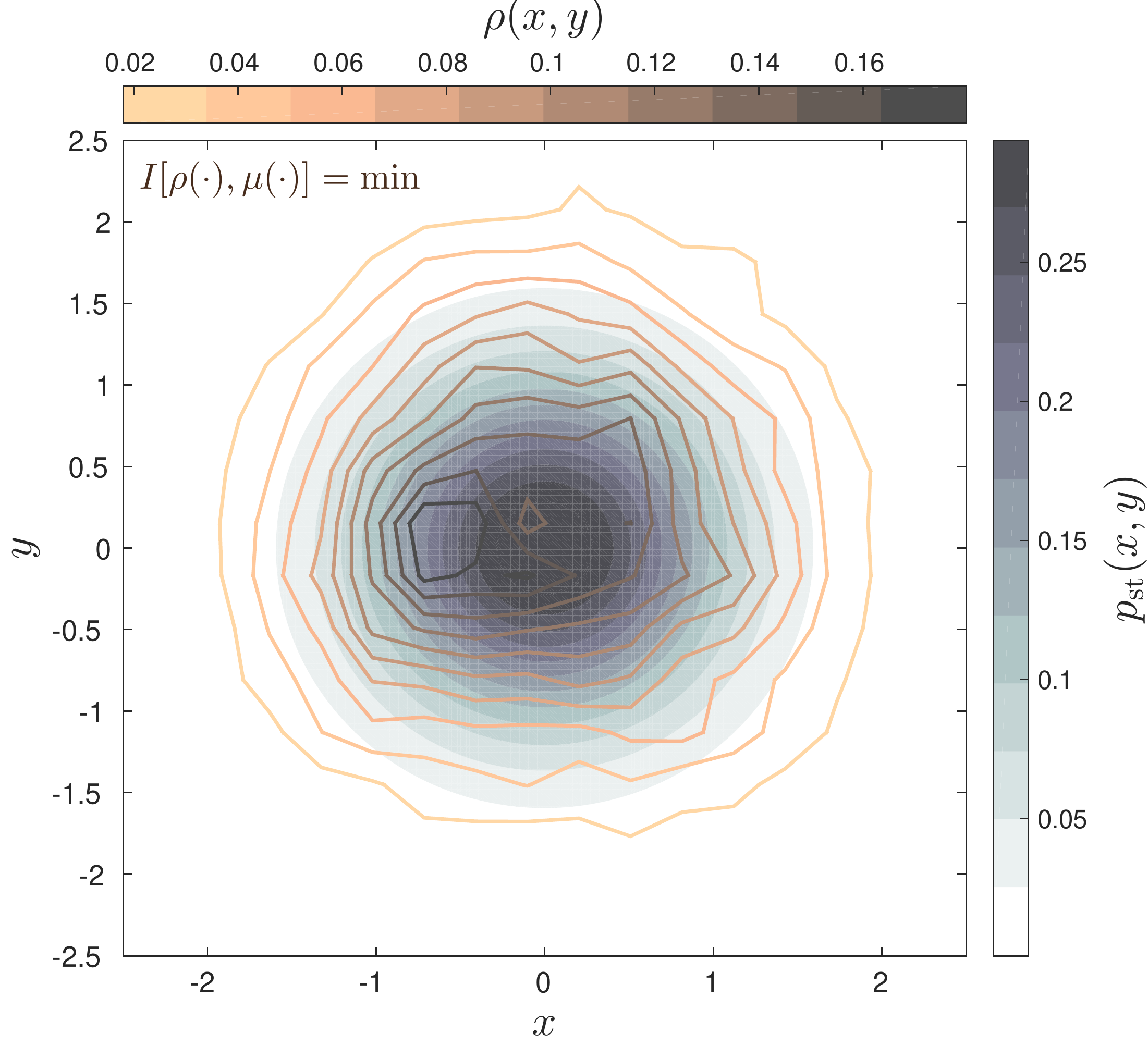}
  \includegraphics[width=0.49\textwidth]{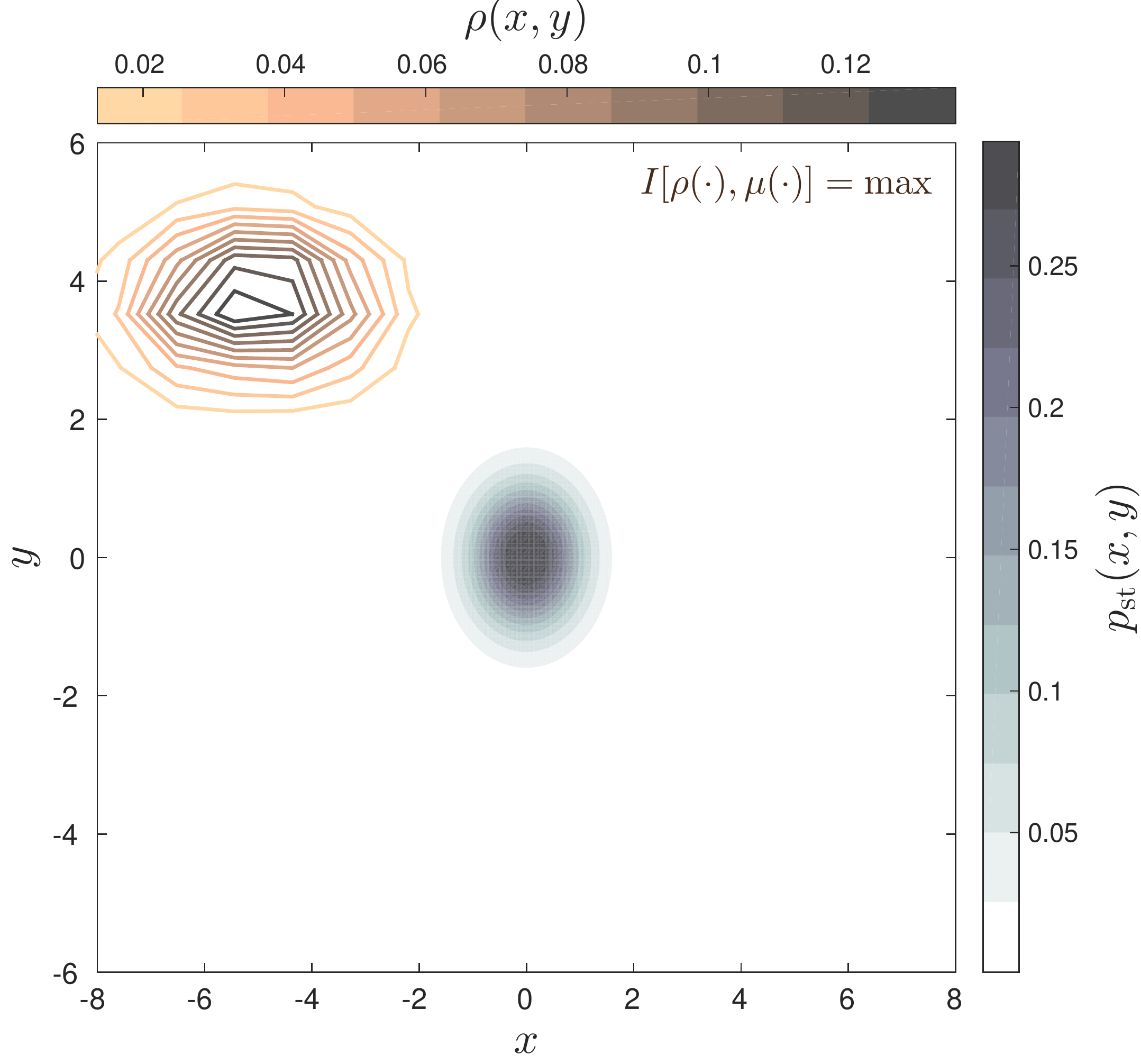}
  \caption{Two realisations of the empirical density $\rho$ that lead to particular low (left) and high (right) values for $I[\rc,\mc)]$. The contours comply with the value of $\rho(x,y)$, the shading represents $\pst(x,y)$. The parameters are $a=1$, $b=2$, $T=500$ and $\beta=2$.}
	\label{fig:wash_rho_sample_minmax}
\end{figure}

\begin{figure}
	\centering
  \includegraphics[width=0.49\textwidth]{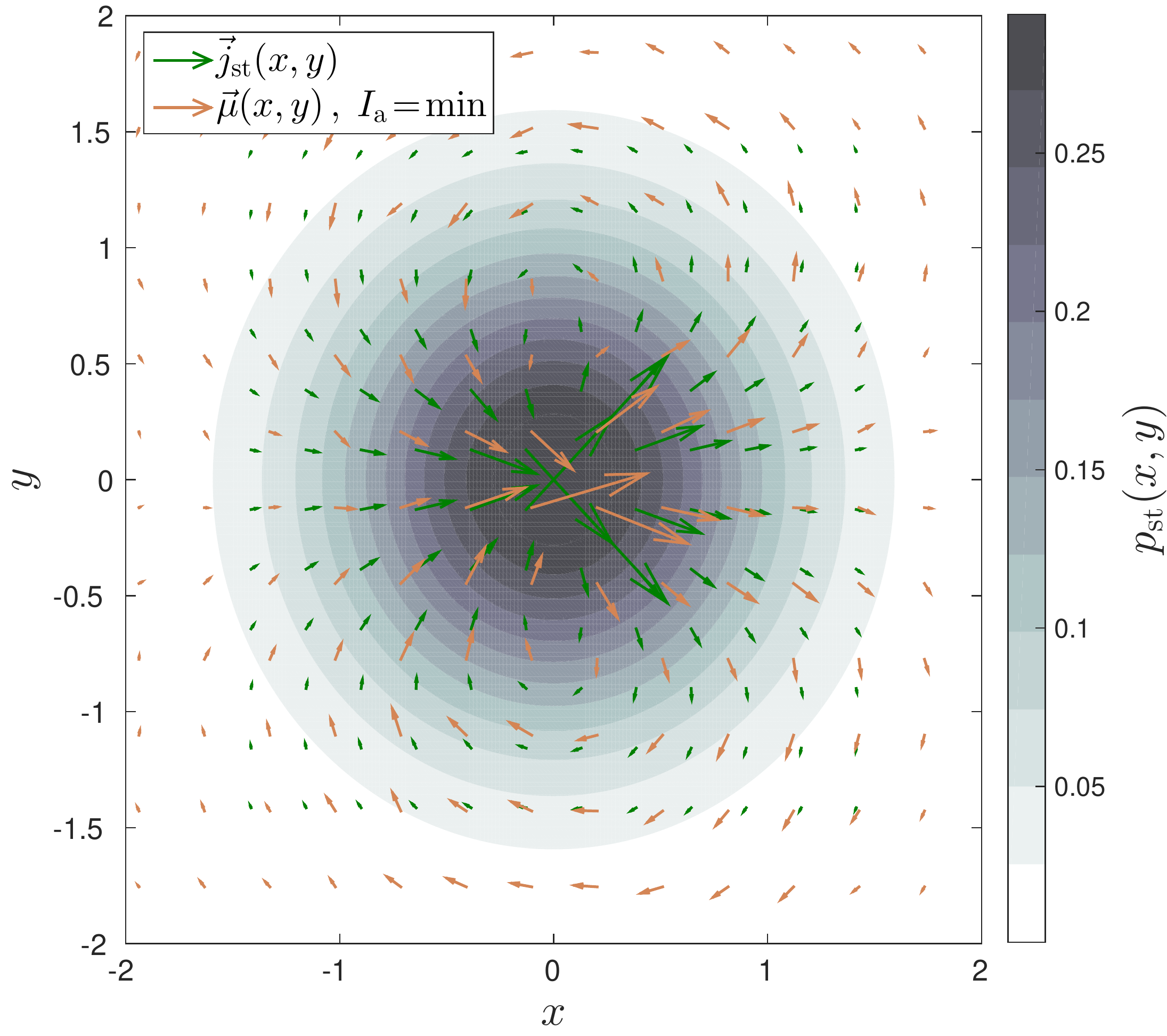}
  \includegraphics[width=0.49\textwidth]{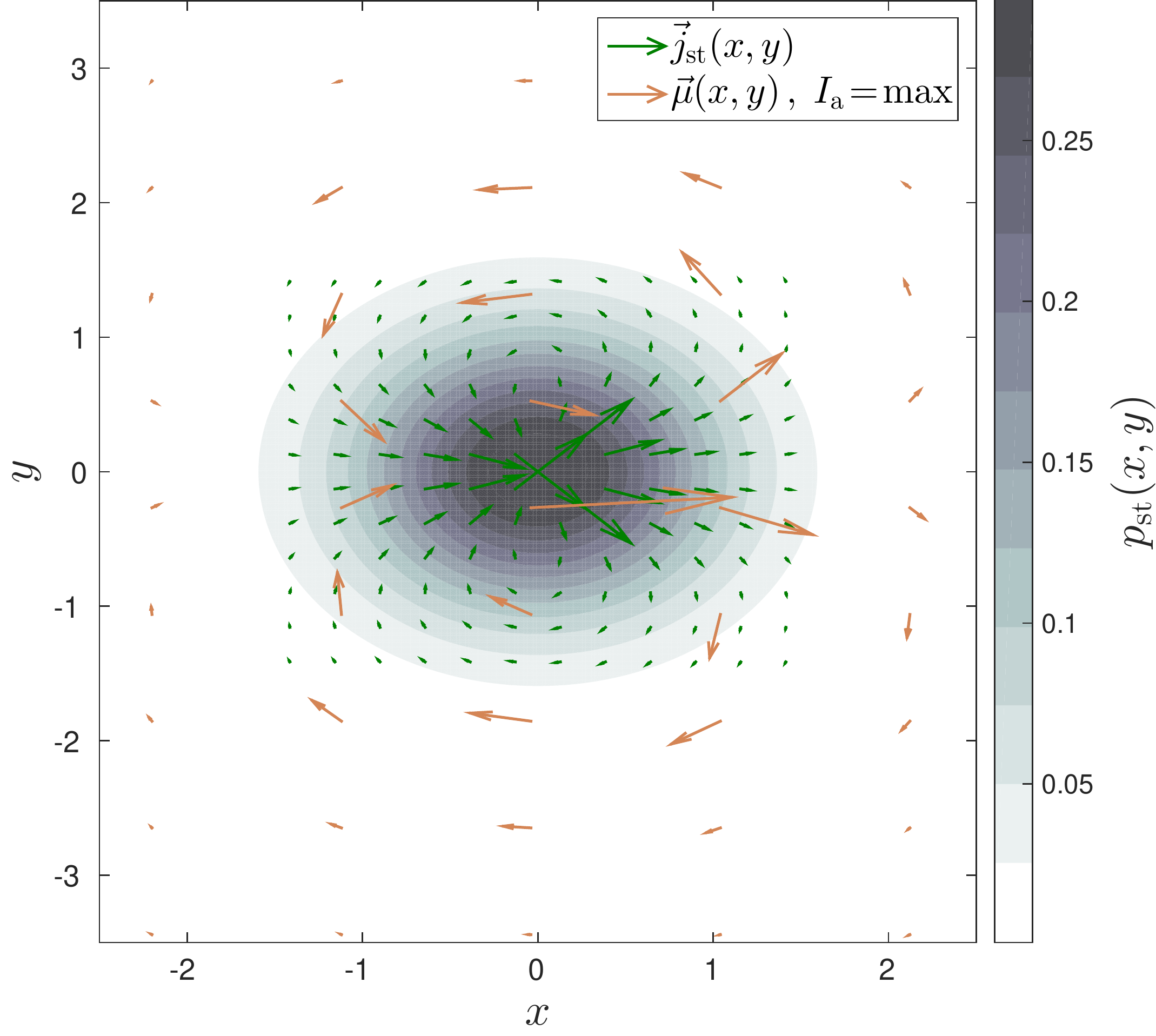}
  \caption{Two realisations of the empirical density $\bmu$ that lead to particular low (left) and high (right) values for $I[\rc,\mc]$. The arrows point in the direction of $\bmu(x,y)$ resp. $\jst(x,y)$, the arrow length complies with $|\bmu(x,y)|$ resp. $|\jst(x,y)|$. The parameters are $a=1$, $b=2$, $T=500$ and $\beta=2$.}
	\label{fig:wash_mu_sample_minmax}
\end{figure}

To consider a numerical example in which fluctuations of $\bmu$ play a role, we keep the stationary distribution \eref{eq:ex1_pst} but alter $\vst$,
\begin{eqnarray}
	\label{eq:ex2_jst}
  \vst(r,\varphi) &= br^3\sin(\varphi)\,\bi e_\varphi + \frac{b}{\beta a}\left(r^2+\frac{2}{\beta a}\right)\frac{\cos(\varphi)}{r}\,\bi e_r
  ,
\end{eqnarray}
ensuring that $\bnabla\cdot\left(\pst(r)\vst(r,\varphi)\right)=0$ still holds. To obtain an ensemble of $\rho$ and $\bmu$ for this system, we repeat the simulation of the Langevin equation with the new force $\bi f(r,\varphi)$ according to \eref{eq:ex1_f} and $\vst(r,\varphi)$ from \eref{eq:ex2_jst}. In \fref{fig:wash_rho_sample_minmax} and \fref{fig:wash_mu_sample_minmax} we show the typical and rare realizations of $\rho$ and $\bmu$. Again, we see that the typical $\rho$ and $\bmu$ are close to $\pst$ and $\jst$ respectively and the rare $\rho$ substantially deviates from $\pst$, but now also the rare $\bmu$ clearly deviates from $\jst$. Hence, for this example, the system fluctuates in both position and direction of movement.

It should furthermore be interesting to perform the contractions discussed in the previous section. Due to the non-linearity of the Euler-Lagrange equations, however, analytical contractions are, even for the simple example \eref{eq:ex1_f}, barely possible. We regard a numerical treatment of the contractions in \sref{sec:marginalization} as beyond the scope of this paper. Instead, we use the parametrization
\numparts
\begin{eqnarray}
  \rho(r) &= \frac{1}{\sqrt{2\pi\sigma^2}}\,\exp\left[ -\frac{r^2}{2\sigma^2} \right] \label{eq:ex_rho}\\
  \bmu(r) &= \rho(r)\,\bnu(r) \,,\quad \bnu(r) = cr \bi e_\varphi \label{eq:ex_mu}
\end{eqnarray}
\endnumparts
to illustrate all findings of the previous section. The empirical density is chosen such that for $\sigma=\sigma_\mathrm{st}\equiv1/\sqrt{\beta a}$ it is $\rho(r)=\pst(r)$, and the empirical current is chosen such that for $c=b$ and $\sigma=\sigma_\mathrm{st}$ it is $\bmu(r)=\jst$ and the constraint $\bnabla\cdot\bmu(r)=0$ is always met.

For this functional choice of $\rho$ and $\bmu$, and $\pst$ and $\jst$ from \eref{eq:ex1_pst} and \eref{eq:ex1_jst}, the rate function $I=\Ia+\Ina$ can be calculated analytically:
\numparts
\begin{eqnarray}  
	\Ia[\sigma,c] &= \frac{\beta\sigma^2}{2}\,(b-c)^2, \label{eq:Ia_sig_c} \\
  \Ina[\sigma] &= \frac{1}{2\beta\sigma^2}\left(\frac{\sigma^2}{\sigma_\mathrm{st}^2} - 1\right)^2. \label{eq:Ina_sig}
\end{eqnarray}
\endnumparts
As expected, it is $I[\sigma=\sigma_\mathrm{st},c=b]=0$.

First, we consider the contraction to $I[\rho]$. In terms of the parametrization \eref{eq:ex_rho}, we are to calculate $I[\sigma]$ by plugging in the optimal $c$ minimizing $I[\sigma,c]$. From $\Ia[\sigma,c]$ above we immediately find that $\bar c(\sigma)\equiv b$ and we are left with the non-adiabatic part 
\begin{equation}
  I[\sigma] = \Ia[\sigma,\bar c\!=\!b]+\Ina[\sigma] = \Ina[\sigma]
  .
\end{equation}

\begin{figure}
	\centering
  \includegraphics[width=0.9\textwidth]{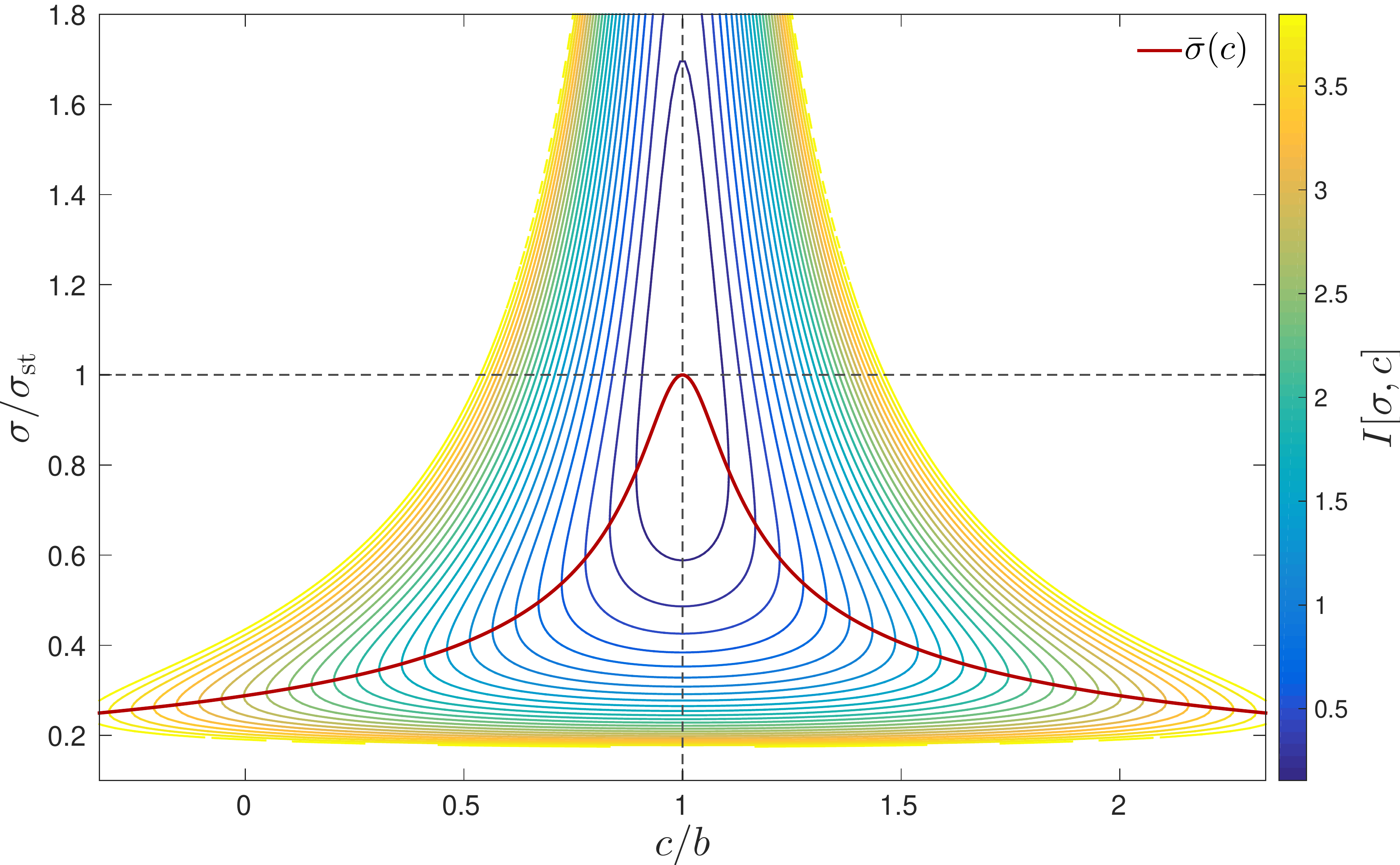}
  \includegraphics[width=0.9\textwidth]{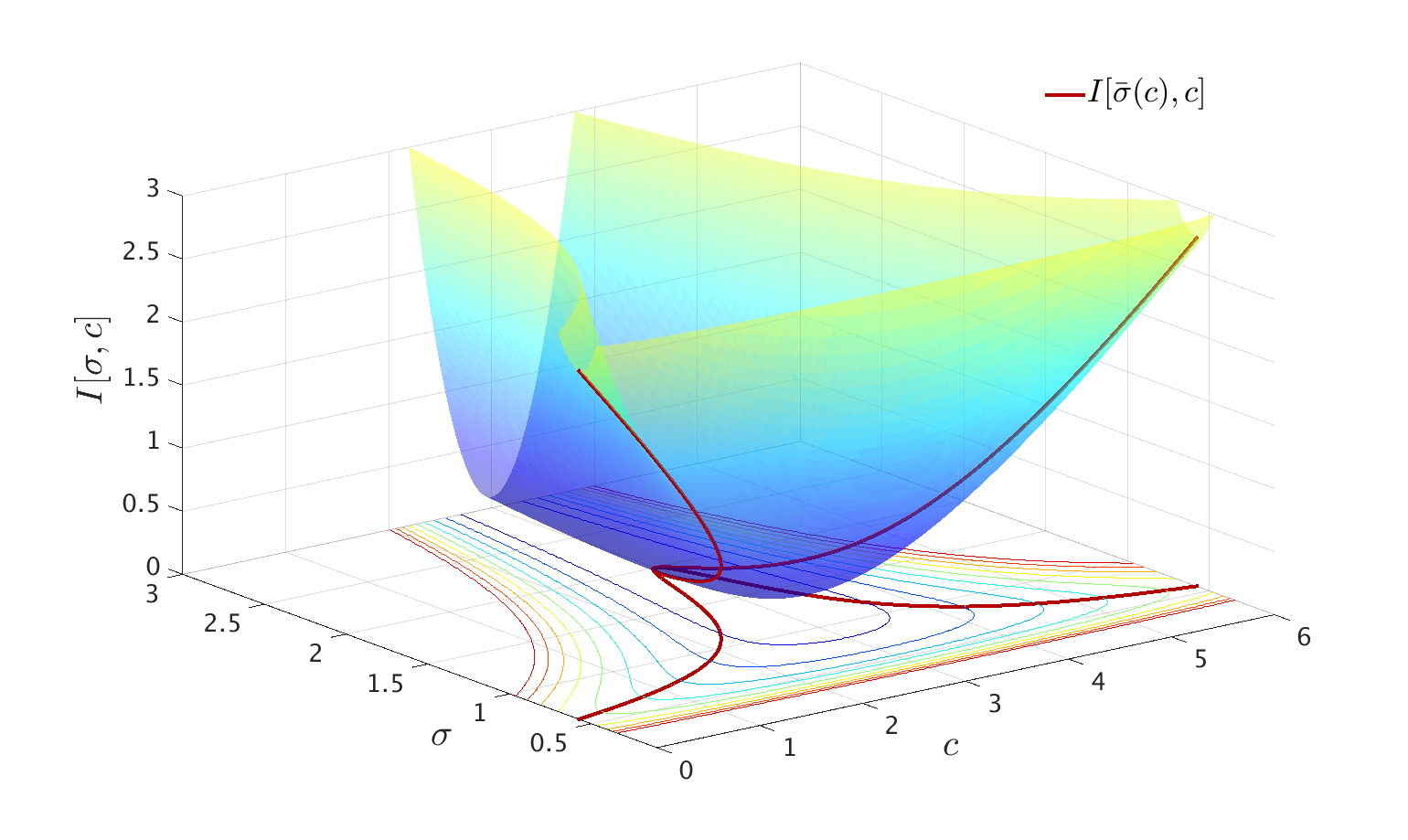}
  \caption{Contour plot and surface plot of $I[\sigma,c]$ with $\sigma=\bar\sigma(c)$ as thick red line.}
	\label{fig:I_c_sigma}
\end{figure}

Next, to illustrate the contraction to $I[\bmu]$, we determine $I[c]$. By minimizing $I[\sigma,c]$ in $\sigma$ we find the optimal $\sigma$ to be
\begin{equation}
	\label{eq:barsig_c}
	\bar\sigma(c)^2 = \frac{\sigma_\mathrm{st}^2}{\sqrt{\beta^2\sigma_\mathrm{st}^4(b-c)^2+1}}
	.
\end{equation}
In figure \ref{fig:I_c_sigma} we plot $I[\sigma,c]=\Ia[\sigma,c]+\Ina[\sigma]$ from \eref{eq:Ia_sig_c} and \eref{eq:Ina_sig} together with the line $\bar\sigma(c)$ from \eref{eq:barsig_c} that minimizes $I[\sigma,c]$ depending on the value of $c$. The dependency of $\bar\sigma(c)$ on $c$ implies that the optimal empirical density is not independent of the empirical current, or, in other words, having sampled an empirical current different from the stationary current, it is likely to also have sampled an empiricial density deviating from the stationary distribution. Thus, we illustrate with this example that we can not just set $I[\rc]=I[\rc,\mc\!\!\equiv\!\!\jst]$ but have to perform the contraction \mbox{$I[\rc]=I[\rc,\mc\equiv\bbmu(\rho)]$} as done in section \ref{sec:marginalization} to derive $I[\rc]$ for systems without detailed balance.
To proceed with the contraction to $I[c]$, we plug $\sigma=\bar\sigma(c)$ in \eref{eq:Ia_sig_c} and \eref{eq:Ina_sig} and get
\begin{eqnarray}
  \Ia[c] &= \Ia[\sigma\!=\!\bar\sigma(c),c] = \frac{(b-c)^2}{2\sqrt{(b-c)^2+1}} ,\\
  \Ina[c] &= \Ina[\sigma\!=\!\bar\sigma(c)] = \frac{\left(\sqrt{(b-c)^2+1}-1\right)^2}{2\sqrt{(b-c)^2+1}} ,\\
  I[c] &= \Ia[c] + \Ina[c] = \sqrt{(b-c)^2+1} - 1 \label{eq:I_c}
  .
\end{eqnarray}
Clearly, both functionals $\Ia[c]$ and $\Ina[c]$ are individually zero for $b=c$.

Finally, we derive by contraction the large deviation function $J(\Sa)$ for this example. From \eref{eq:Sa_mu} follows that
\begin{equation}
  \Sa(b) = 2\beta T\sigma^2cb 
  .
\end{equation}
Substitution into \eref{eq:Ia_sig_c} gives in agreement with the first term in \eref{eq:Ia_rho_Sa}
\begin{eqnarray}
  \Ia[\sigma,\Sa] &= \frac{\left(2\beta T\sigma^2b^2-\Sa\right)^2}{8\beta T^2\sigma^2b^2} \nonumber\\
  &= \frac{\left(\dSa(\sigma)-\frac{\Sa}{T}\right)^2}{4\dSa(\sigma)}
  ,
\end{eqnarray}
where in the second line we plugged in $\dSa(\sigma) = 2\beta\sigma^2b^2$ from \eref{eq:Sa}. 
In the last step we optimize $I[\sigma,\Sa] = \Ia[\sigma,\Sa] + \Ina[\sigma]$ in $\sigma$,
\begin{equation}
	\bar\sigma(\Sa)^2 = \sigma_\mathrm{st}^2 \sqrt{\frac{\left(\frac{\Sa}{T}\right)^2+4b^2}{\dSa^2+4b^2}}
	,
\end{equation}
and find by substitution into $I[\sigma,\Sa]$ the desired large deviation function
\begin{eqnarray}
  J(\Sa) &= I[\bar\sigma(\Sa),\Sa] \nonumber\\
  &= \frac{1}{2}\left(\dSa+\frac{4b^2}{\dSa}\right)\sqrt{\frac{\left(\frac{\Sa}{T}\right)^2+4b^2}{\dSa^2+4b^2}} - \frac{1}{2}\left(\frac{\Sa}{T}+\frac{4b^2}{\dSa}\right)
  .
\end{eqnarray}
As expected, it is $J(\Sa\!=\!T\dSa)=0$, and we see that $p(\Sa)$ has exponential tails.



%% file: appendix.tex
\appendix
\section{Degenerate Eigenvalues}
\label{sec:appendix_DE}
In general the operator $L$ is not hermitian and it is hence not guaranteed that $L$ has a complete set of eigenfunctions.
Here, we argue that the result that $\lambda[q(\cdot)]$ can be identified with the negative of the largest eigenvalue of the operator $L_{q(\cdot)}$ (and $L_{\bm \gamma(\cdot),\eta(\cdot)}$ respectively) also holds if $L$ is not diagonalizable. For notational simplicity we omit the dependency of $L$ on $q(\cdot)$, $\bm \gamma(\cdot)$ and $\eta(\cdot)$ and use the Einstein summation convention. 

We assume, that $G(\bm x, t| \bm x_0, 0)$ can be written as linar combination of $N$ orthogonal functions $f_n(\bm x)$:
\begin{equation}
  G(\bm x, t | \bm x_0, 0) = g_n(t | \bm x_0, 0) f_n(\bm x)
\end{equation}
with 
\begin{equation}
  g_n(t | \bm x_0, 0) = \int \mathrm d \bm x \, f_n^*(\bm x) G(\bm x, 0| \bm x_0, 0).
\end{equation}
Using this form of $G$, the differential equations \eref{eq:G_pde} and \eref{eq:pde_ness} turn into 
\begin{equation}
  \partial_t g_m(t) = \bar L_{m,n} g_n(t)
\end{equation}
with the matrix
\begin{equation}
  \bar L_{m,n} = \int \mathrm d \bm x \, f_m^*(\bm x) L f_n(\bm x).
\end{equation}
The formal solution of this differential equation is given by
\begin{equation}
  g_n(T| \bm x_0, 0) = \left( e^{T \bar L}\right)_{n,m} g_m(0| \bm x_0, 0)
\end{equation}
with the initial condition $g_n(0| \bm x_0, 0) = \int \mathrm d \bm x \, f_n^*(\bm x) \delta(\bm x - \bm x_0) = f_n^*( \bm x_0)$.
Inserting this into \eref{eq:Q_pathintegral} or \eref{eq:Q_pathintegral_ness} gives for the cumulant generating function 
\begin{equation}
  \label{eq_Appendix_G_sol}
  Q_T = \int \mathrm d \bm x_T \, \int \mathrm d \bm x_0 \, p(x_0)
  f_n(\bm x_T) \left( e^{T \bar L}\right)_{n,m} f^*_m(x_0).
\end{equation}
Let us now introduce a similarity transformation $\bar L = H J H^{-1}$. 
For the matrix exponential, this leads to $e^{T \bar L} = H e^{TJ} H^{-1}$.

If $\bar L$ has $N$ different eigenvalues $\lambda_n$, we chose the columns of $H$ to be the eigenvectors of $\bar L$. Then $J$ is a diagonal matrix with the eigenvalues  on the diagonal. This leads to $(e^{T  J })_{m,n} = \delta_{m,n} e^{T \lambda_n}$. If we put this into \eref{eq_Appendix_G_sol} and further identify $\phi^\nu = f_m(x)H_{m,\nu}$ and $\psi^\nu = H^{-1}_{\nu,n}f^*_n(x)$ we recover \eref{eq:Q_sol} and \eref{eq:Q_sol_ness} respectively.

If $\bar L$ has only $k < N$ different  eigenvalues $\lambda_n$, that are of algebraic multiplicity $s_n$, we chose the columns $\bm y_{(1,1)}, \ldots,\bm y_{(1, s_1)}, \ldots,  \bm y_{(k,1)}, \ldots,\bm y_{(k,s_k)}$ of $H$ to be the generalized eigenvectors of $\bar L$, i.e. $\bm y_{(n,i)} \in \ker (\bar L - \lambda_n \mathrm{id})^i $. In analogy, we denote the rows of $H^{-1}$ by $\bm z_{(1,1)}, \ldots,\bm z_{(1, s_1)}, \ldots,  \bm z_{(k,1)}, \ldots,\bm z_{(k,s_k)}$. This choice of $H$ allows $J$ to be of Jordan normal form:
\begin{equation}
  J = \pmatrix{
    J_1 & \, & \, \cr
    \, & \ddots & \, \cr
    \, & \, & J_k
  },
\end{equation}
where the Jordan blocks $J_n$ are $s_n \times s_n$ matrices, with $\lambda_n$ on the diagonals and ones on the superdiagonals. For the exponential, this gives 
\begin{equation}
  e^{tJ} = 
  \pmatrix{
    e^{t J_1} & \, & \, \cr
    \, & \ddots & \, \cr
    \, & \, & e^{t J_k}
  },
\end{equation}
where the block matrices are given by 
\begin{equation}
  \left(e^{t J_n}\right)_{i,j} = \cases{
  0 & if $i > j$, \cr
   e^{t \lambda_n} \frac{t^{j-i}}{(j-i)!} & otherwise.}
\end{equation}
Inserting all this into \eref{eq_Appendix_G_sol} gives 
\begin{equation}
  Q_T = \sum_{n=1}^k e^{T \lambda_n}
  \int \rmd \bm x_T \, \int \rmd \bm x_0 \, p_0(\bm x_0) \,
    \sum_{m=1}^{s_n} \sum_{l=0}^{s_n - m}
    \phi_{(n,m)}(\bm x_T) \frac{T^{l}}{l!} \psi_{(n,m+l)}(\bm x_0),
\end{equation}
with $\phi_{(n,m)} = f_j(\bm x_T) y_{j,(n,m)}$ and $\psi_{(n,m+l)} = z_{(n,m+l),j} f^*_j(\bm x_0)$.
Hence, like in the case with $N$ different eigenvalues, for large $t$, $Q_T$ is dominated by $e^{\lambda_1 t}$, where $\lambda_1$ is the eigenvalue with the largest real part and the integrals over $\bm x_T$ and $\bm x_0$ are part of the $\mathrm o(T)$ therm in \eref{eq:lambda_def1} and \eref{eq:lambda_def_ness} .


